# Friedmann limits of rotating hypersurface-homogeneous dust models

Andrzej Krasiński[*]

N. Copernicus Astronomical Center, Polish Academy of Sciences
Bartycka 18, 00 716 Warszawa, Poland
and School of Sciences, Polish Academy of Sciences
email: akr@camk.edu.pl

**Abstract**

The existence of Friedmann limits is systematically investigated for all the hypersurface-homogeneous rotating dust models, presented in previous papers by this author. Limiting transitions that involve a change of the Bianchi type are included. Except for stationary models that obviously do not allow it, the Friedmann limit expected for a given Bianchi type exists in all cases. Each of the 3 Friedmann models has parents in the rotating class; the $k = +1$ model has just one parent class, the other two each have several parent classes. The type IX class is the one investigated in 1951 by Gödel. For each model, the consecutive limits of zero rotation, zero tilt, zero shear and spatial isotropy are explicitly calculated.

## I. Motivation and summary of the method.

In previous papers[1–3] a complete set of all metric forms was derived that can represent hypersurface-homogeneous rotating dust models. For each case, the generators of the symmetry algebra were found, the Bianchi type determined, and the metric form resulting from the Killing equations was explicitly presented. That classification was more detailed than the Bianchi classification because all possible orientations of the symmetry orbits in the spacetime were allowed, i.e. the orbits could be spacelike, timelike or null.

In a later paper[4], one of the Bianchi type V models was investigated. Among the problems considered there was the question whether the model can reproduce the $k = -1$ Friedmann model in the limit of zero rotation, $\omega \to 0$. Since the coordinates that are well-suited to the classification are not suitable at all for considering the limit $\omega \to 0$, this limit could be taken only after a coordinate change and reparametrization of the metric.

In the present paper, the existence of the Friedmann limits is systematically investigated for all the other cases found in the classification in Refs. 1–3. The Bianchi type is allowed to change in the limiting transition. In all Bianchi type I cases the velocity field is tangent to the symmetry orbits, i.e. those models have matter density constant along the flow, and no expanding Friedmann model can be a subcase there. The same is true for the Bianchi type II from Ref. 1 and for both the subcases of case 1.1.1.2 in Ref. 2 which are of type III. In all the other cases the Friedmann limits that can be expected for a given Bianchi type do indeed exist.

---

[*]This research was supported by the Polish Research Committee grant no 2 P03B 060 17.

The specialization to the Friedmann metrics is possible in so many cases because there is a free parameter in them that determines the tilt of the orbits with respect to the velocity field (with various values of the tilt parameter, the orbits may be spacelike, timelike or null). Whenever a Friedmann limit exists, the orbits are made orthogonal to the velocity field ("untilted") during the limiting transition.

In order to make this paper readable independently of the other ones, the basic facts are briefly recalled here. More details can be found in Ref. 1.

The velocity field of a rotating dust, $u^\alpha$, defines 3 scalar functions $\tau(x), \eta(x)$ and $\xi(x)$ such that:

$$u_\alpha = \tau_{,\alpha} + \eta \xi_{,\alpha}. \tag{1.1}$$

These functions (whose existence follows from the equations of motion via the Darboux theorem[1]) are determined up to the transformations:

$$\tau = \tau' - S(\xi', \eta'), \qquad \xi = F(\xi', \eta'), \qquad \eta = G(\xi', \eta'), \tag{1.2}$$

where the functions $F$ and $G$ obey:

$$F_{,\xi'} G_{,\eta'} - F_{,\eta'} G_{,\xi'} = 1, \tag{1.3}$$

and then $S$ is determined by:

$$S_{,\xi'} = GF_{,\xi'} - \eta', \qquad S_{,\eta'} = GF_{,\eta'}. \tag{1.4}$$

(eq. (1.3) is the integrability condition of (1.4)).

Then, the continuity equation, $(nu^\alpha)_{;\alpha} = 0$, where $n$ is the number density of the dust particles, implies that there exists one more function $\zeta(x)$ such that:

$$\sqrt{-g} n u^\alpha = \varepsilon^{\alpha\beta\gamma\delta} \xi_{,\beta} \eta_{,\gamma} \zeta_{,\delta}, \tag{1.5}$$

where $g$ is the determinant of the metric tensor and $\varepsilon^{\alpha\beta\gamma\delta}$ is the Levi-Civita symbol. The function $\zeta$ is determined up to the transformations:

$$\zeta = \zeta' + T(\xi', \eta'), \tag{1.6}$$

where $T$ is an arbitrary function.

The following relations hold then:

$$u^\alpha \tau_{,\alpha} = 1, \qquad u^\beta \xi_{,\beta} = u^\beta \eta_{,\beta} = u^\beta \zeta_{,\beta} = 0.$$

$$\frac{\partial(\tau, \eta, \xi, \zeta)}{\partial(x^0, x^1, x^2, x^3)} = \sqrt{-g} n \neq 0. \tag{1.7}$$

This shows that $\{\tau, \xi, \eta, \zeta\}$ can be chosen as coordinates, with $\tau$ being the time coordinate. They are called the Plebański coordinates. Denoting $\{\tau, \xi, \eta, \zeta\} = \{x^0, x^1, x^2, x^3\} =$ $\{t, x, y, z\}$, we obtain for the velocity field $u^\alpha$, the metric tensor $g_{\alpha\beta}$, the rotation tensor $\omega_{\alpha\beta}$ and the rotation vector $w^\alpha$ in these coordinates:

$$u_\alpha = \delta^\alpha{}_0, \qquad u_\alpha = \delta^0{}_\alpha + y\delta^1{}_\alpha,$$

$$g_{00} = 1, \qquad g_{01} = y, \qquad g_{02} = g_{03} = 0, \qquad g \equiv \det(g_{\alpha\beta}) = -n^{-2},$$

$$w^\alpha = n\delta^\alpha_3, \qquad \omega_{\alpha\beta} = -\omega_{\beta\alpha} = \frac{1}{2}\delta^1{}_\alpha \delta^2{}_\beta. \tag{1.8}$$

It is the last property that makes the limitng transition $\omega \to 0$ impossible without a coordinate transformation and reparametrization.

In these coordinates, if any Killing field is allowed by the metric it must be of the form:

$$k^\alpha = (C + \phi - y\phi_{,y})\delta^\alpha{}_0 + \phi_{,y}\,\delta^\alpha{}_1 - \phi_{,x}\,\delta^\alpha{}_2 + \lambda\delta^\alpha{}_3, \tag{1.9}$$

where $C$ is an arbitrary constant and $\phi(x, y)$ and $\lambda(x, y)$ are arbitrary functions. If $\phi_{,\alpha} \neq 0$ (i.e. $\phi$ is not constant on an open set), then the coordinates can be adapted to $k^\alpha$ within the Plebański class (by eqs. (1.2) – (1.4) and (1.6)) so that:

$$k^\alpha = \delta^\alpha{}_1. \tag{1.10}$$

The metric then becomes independent of $x$, and the coordinates preserving (1.10) are determined up to the transformations:

$$t' = t - \int yH_{,y}\, dy + A, \qquad x' = x + H(y), \qquad y' = y, \qquad z' = z + T(y), \tag{1.11}$$

where $A$ is an arbitrary constant and $H, T$ are arbitrary functions.

If $\phi_{,\alpha} = 0$, then the form of the Killing field $k^\alpha = C\delta^\alpha{}_0 + \lambda\delta^\alpha{}_3$, is invariant under (1.2) – (1.6) and the Plebański coordinates cannot be adapted to $k^\alpha$. The property $\phi = \text{const}$ is equivalent to the following invariant relation:

$$k^\alpha = Cu^\alpha + (\lambda/n)w^\alpha, \tag{1.12}$$

i.e. $k^\alpha$ is then spanned on the velocity field and the rotation field.

If three Killing fields exist, then each of them can either be of the special type (1.12) or of the general type (1.9). One of the general-type Killing fields can always be transformed to the form (1.10) by (1.2) – (1.6).

This observation gives rise to a complete classification of all hypersurface-homogeneous spacetimes that are possible for a rotating dust. When all 3 Killing fields are of the special type (1.12), the symmetry orbits are two-dimensional, and this case is not considered. When two Killing fields are of the special type, while the third one is general, there exist two classes of metrics (Bianchi types I and II) that were derived in Ref. 1. When one Killing field is of the special type, while the two others are general, all Bianchi types except VIII and IX appear (Ref. 2). When all 3 Killing fields are of the general type, all the Bianchi types appear, some of them hidden as limits of more general types (Ref. 3).

The multitude of cases is a consequence of the many possible alignments or misalignments among the 3 Killing fields and the velocity and rotation fields.

When the Bianchi classification is introduced, the generators of symmetry are scaled to standard forms such that all nonzero structure constants (except the free parameters in types $VI_h$ and $VII_h$) become equal either to $+1$ or to $-1$. In general, though, they are arbitrary constants, and in the general form each of those constants can be allowed to become zero. In this way, the more special Bianchi types can be obtained from the more general ones by going to the zero limit with some of the structure constants. The resulting hierarchy of Bianchi types is well-known, and is shown in Fig. 1 (adapted from Ref. 5) for easy reference. The specializations that are possible can be instantly guessed from the values of the $a, n_1, n_2$ and $n_3$ parameters for the different Bianchi types. Type III cannot be specialized to IV or V because, with the arbitrary values of the parameters $n_2$ and $n_3$, the parameter $a$ is determined by $a = \sqrt{-n_2 n_3}$.

Another well-known result[6] is the placement of different Robertson-Walker geometries within the Bianchi classes. This is also recalled for easy reference. Since we are considering only dust models, we will call these geometries the Friedmann models and Friedmann limits of the rotating models.

The $k = 0$ model is a subcase of the Bianchi types I and $VII_0$ (the two Bianchi algebras have different bases, but share common orbits).

The $k = -1$ model is a subcase of the Bianchi types V and $VII_h$.

The $k = +1$ model is a subcase of the Bianchi type IX.

When considering each case of the classification from Refs. 1–3, one has to recognize from Fig. 1 which of the four types $\{I, V, VII_0, VII_h, IX\}$ could possibly be contained in it as a subcase and then the appropriate specialization of the arbitrary constants and functions in the model has to be considered. This procedure will be presented in more detail in sec. II, later it will be applied without detailed explanations.

It will turn out that only the stationary models have no Friedmann limit. In every nonstationary case, the Friedmann limit indicated by the Bianchi type indeed exists. Note that the limits are found for the metrics, without taking into account the Einstein equations. This is why a nonstationary type II metric exists in the collection, and is found to admit the $k = 0$ Friedmann limit, even though it is known [7,8] that spatially homogeneous type II dust solutions must have zero rotation, see sec. III.

Now we shall systematically go over all the cases presented in Refs. 2 and 3. The two cases from Ref. 1 are immediately seen to admit no Friedmann limit: in both of them, the velocity field of the dust is spanned on the Killing fields (see eqs. (7.7) – (7.8) is Ref. 1), so the particle number density $n$ will obey $n_{,\alpha} u^\alpha = 0$. Hence, these cases cannot contain any expanding Firedmann model because in the latter $n_{,\alpha} u^\alpha \neq 0$.

Each of the models presented in Refs. 2 and 3 that allows a Friedmann limit will be first transformed to the Plebański coordinates (most of them were found in coordinates adapted to the Killing fields that are not in the Plebański class). Then, each model will be transformed by a coordinate transformation and reparametrization of the metric functions and constants to such a form in which the limit of zero rotation can be calculated explicitly. Then, the Friedmann limits will be calculated by consecutively imposing on the metric the conditions of zero rotation, zero tilt, zero shear and spatial isotropy (i.e. constant curvature in the 3-spaces orthogonal to the dust flow). This last condition is not superfluous, even though dust with zero rotation and zero shear must be a Friedmann model in consequence of the Einstein equations[9]. It is conceivable that no Friedmann limit would exist at all in some classes. However, this does not happen, and a spatially isotropic subcase will be found to exist in all cases. The corresponding limits of the Killing fields, where nontrivial, will be also calculated and the Bianchi type of the limit determined.

Since on each of the underlying manifolds five vector fields exist (velocity, rotation and the 3 Killing fields), the 5 vectors must be linearly dependent at each point. This linear relation allows to identify in each case the parameter that determines the tilt of the velocity field with respect to the symmetry orbits - see sec. V. It turns out that this tilt parameter is always simply proportional to that defined by King and Ellis[8].

## II. The cases 1.1.1 of Ref. 2.

We begin with case 1.1.1.1., which is of Bianchi type III.

The transformation from the coordinates used in eq. (2.18) of Ref. 2 (that were adapted to the Killing fields) to the Plebański coordinates is given by eq. (2.16) in Ref. 2 (where $\{t', x', Y, Z\}$ are the coordinates of (2.18), and $\{t, x, y, z\}$ are the Plebański coordinates). The transformed metric is:

$$g_{00} = 1, \qquad g_{01} = y, \qquad g_{02} = g_{03} = 0,$$

$$g_{11} = \left(\frac{Y}{2\lambda_3}\right)^2 + \frac{1}{2\lambda_3^2} YZ + h_{11} - \frac{b}{\lambda_3} h_{12} Y + \left(\frac{1}{2} bY\right)^2 h_{22},$$

$$g_{12} = h_{12} - \frac{1}{2} b\lambda_3 Y h_{22}, \qquad g_{13} = C_3 h_{13} - \frac{1}{2} b^2 C_3 Y h_{23},$$

$$g_{22} = \lambda_3^2 h_{22}, \qquad g_{23} = bC_3 \lambda_3 h_{23}, \qquad g_{33} = (bC_3)^2 h_{33}, \tag{2.1}$$

where $b$, $C_3$ and $\lambda_3$ are arbitrary constants, $Y$ and $Z$ are given by:

$$Y = -b\lambda_3 t + \lambda_3 y + bC_3 z, \qquad Z = b\lambda_3 t + \lambda_3 y - bC_3 z, \tag{2.2}$$

and all the $h_{ij}$, $i, j = 1, 2, 3$ are arbitrary functions of $Z$. The first line of eq. (2.1) will be the same in all the other metrics transformed to the Plebański coordinates, so it will not be repeated from now on. Since the argument of $h_{ij}$ is determined (by the Killing equations) only up to a constant factor, we are allowed to rescale it by an arbitrary factor. For considering the limit $\omega \to 0$, it will be convenient to assume that the argument of $h_{ij}$ is:

$$T := Z/(b\lambda_3) = t + y/b - (C_3/\lambda_3)z. \tag{2.3}$$

This presupposes that $b\lambda_3 \neq 0$, but this condition is included in the definition of case 1.1.1.1. The limit $\lambda_3 = 0$ can be taken into account after a simple reparametrization, and it leads to a stationary solution. The subcase $b = 0$ is degenerate, and it belongs to the 1.1.2 family.

As seen from the last formula in (1.8), the simplest way to let $\omega \to 0$ is to transform $y$ as follows:

$$y = \omega_0 \tilde{y}, \tag{2.4}$$

and then let $\omega_0 \to 0$, so that the only nonzero component of rotation in the new coordinates becomes:

$$\omega'_{12} = \omega_0 \tilde{y} \xrightarrow[\omega_0 \to 0]{} 0. \tag{2.5}$$

Then, however, the components $g'_{12}$, $g'_{22}$ and $g'_{23}$ of the transformed metric would simultaneously go to zero, and the metric would become degenerate ($g = 0$). To avoid this, $h_{22}$ must be rescaled as follows:

$$h_{22} = H_{22}/\omega_0^2. \tag{2.6}$$

Then $g'_{12} = \omega_0 h_{12} - \frac{1}{2} b \lambda_3 Y H_{22}/\omega_0$ would become infinite in the limit $\omega_0 \to 0$. To avoid this, $h_{12}$ must be reset so that the infinity is cancelled. Since all $h_{ij}$ depend on $T$, not on $Y$, this can be done as follows:

$$h_{12} = H_{12}/\omega_0 - \frac{1}{2}(b\lambda_3)^2 T H_{22}/\omega_0^2. \tag{2.7}$$

The first term in (2.7) contains the $\omega_0$ in the denominator for greater generality, so that $g'_{12} \xrightarrow[\omega_0 \to 0]{} H_{12} \neq 0$. Then, to cancel the infinities in $g_{11}$, the function $h_{11}$ must be reset as follows:

$$h_{11} = H_{11} - \frac{1}{4} b^4 (\lambda_3 T)^2 H_{22}/\omega_0^2 - b^2 T h_{12}. \tag{2.8}$$

The reparametrization (2.4), (2.6) – (2.8) would be sufficient to make the limit $\omega_0 \to 0$ of the metric (2.1) nondegenerate. However, the hypersurfaces $t = $ const, that become orthogonal to the velocity field $u^\alpha$ in the limit $\omega_0 \to 0$, would not yet coincide with the hypersurfaces of constant matter density. In the Plebański coordinates, as seen from (1.8), the matter density obeys $g = -n^{-2}$, and so $n$ would depend on $(t - C_3 z/\lambda_3)$ in the limit $\omega_0 \to 0$, i.e. the model would still be tilted. To "untilt" it, we must let $C_3 \to 0$, and this requires at least one more rescaling. It will be convenient to redefine $C_3$ as follows:

$$C_3 = \omega_0 c, \tag{2.9}$$

so that the untilting occurs simultaneously with $\omega \to 0$. Then we must rescale $h_{33}$:

$$h_{33} = H_{33}/\omega_0^2. \tag{2.10}$$

For greater generality, we will also rescale $h_{23}$:

$$h_{23} = H_{23}/\omega_0^2, \tag{2.11}$$

and then $h_{13}$ must be reset as follows:

$$h_{13} = H_{13}/\omega_0 - \frac{1}{2} b^3 \lambda_3 T H_{23}/\omega_0^2. \tag{2.12}$$

The transformation (2.4), applied to (2.1) together with all the subsequent reparametrizations, results in the following metric:

$$g_{00} = 1, \quad g_{01} = \omega_0 \tilde{y}, \quad g_{02} = g_{03} = 0,$$

$$g_{11} = \frac{1}{4\lambda_3{}^2}[b\lambda_3 t(-2\lambda_3 \tilde{y} + 2bcz)\omega_0 + (\lambda_3 \tilde{y} + bcz)(3\lambda_3 \tilde{y} - bcz)\omega_0{}^2]$$

$$-\frac{1}{4}(bt)^2 + H_{11} - 2b\tilde{y}H_{12} + (b\lambda_3 \tilde{y})^2 H_{22},$$

$$g_{12} = H_{12} - b\lambda_3{}^2 \tilde{y} H_{22}, \quad g_{13} = cH_{13} - b^2 c\lambda_3 \tilde{y} H_{23},$$

$$g_{22} = \lambda_3{}^2 H_{22}, \quad g_{23} = bc\lambda_3 H_{23}, \quad g_{33} = (bc)^2 H_{33}, \tag{2.13}$$

where the $H_{ij}$ depend only on $t$. Here, similarly as in (2.1), the first line will be the same for every metric, and so it will not be repeated from now on.

The metric (2.13) still has nonzero shear. If a Friedmann model is to result from it, the shear must be set to zero. The coordinates $\{t, x, y, z\}$ in (2.13) are now comoving and synchronous, so zero shear means that:

$$g_{ij} = G_{ij}(x, y, z) R^2(t), \tag{2.14}$$

i.e. all the components of the metric must depend on time only through the same factor $R^2(t)$. This means:

$$H_{11}(t) = \frac{1}{4}b^2 t^2 - C_{11} R^2(t)$$
$$\text{other } H_{ij}(t) = -C_{ij} R^2(t), \tag{2.15}$$

where $C_{ij}$ are unknown constants. With no loss of generality, it may be assumed that:

$$C_{33} = 1. \tag{2.16}$$

The metric (2.13) with $H_{ij}$ as in (2.15) – (2.16) will represent a Friedmann model when the hypersurfaces $t = $ const are spaces of constant curvature. In order to calculate this curvature, it is convenient to introduce the new constants $D_{11}$, $D_{12}$ and $D_{22}$ by:

$$D_{22}{}^2 := C_{22} - C_{23}{}^2, \quad D_{12} := (C_{12} - C_{13} C_{23}/b)/(\lambda_3 D_{22}),$$

$$D_{11}{}^2 := C_{11} - C_{13}{}^2/b^2 - D_{12}{}^2. \tag{2.17}$$

The correct signs for $D_{11}{}^2$ and $D_{22}{}^2$ are guaranteed by the signature of (2.13). Then (2.13) may be written as follows:

$$\mathrm{d}s^2 = \mathrm{d}t^2 - (D_{11} R \mathrm{d}x)^2 - R^2[(D_{12} - b\lambda_3 D_{22} y)\mathrm{d}x + \lambda_3 D_{22} \mathrm{d}y]^2$$

$$- R^2[(C_{13}/b - b\lambda_3 C_{23} y)\mathrm{d}x + \lambda_3 C_{23} \mathrm{d}y + bc \mathrm{d}z]^2, \tag{2.18}$$

and the curvature tensor for the spaces $t = $ const may be easily calculated using the orthonormal set of differential forms suggested by (2.18). The curvature tensor is:

$$R^{12}{}_{12} = \frac{3}{4} F^2 G^2 + F^2, \quad R^{13}{}_{13} = R^{23}{}_{23} = -\frac{1}{4} F^2 G^2, \tag{2.19}$$

where:
$$F := b/(D_{11}R), \qquad G := C_{23}/D_{22}. \tag{2.20}$$

The Riemann tensor (2.19) will represent constant curvature when $R^{12}{}_{12} = R^{13}{}_{13}$. This implies $b = 0$, which seems to be a singular limit of (2.18). However, the limit $b \to 0$ may be easily incorporated into (2.18) by the following reparametrization:

$$C_{13} = D_{13}b, \qquad c = C/b. \tag{2.21}$$

After this, the Riemann tensor of the space $t = $ const still has the same form (2.19) – (2.20). With $b = 0$, $R^{ij}{}_{kl} \equiv 0$, i.e. (2.18) represents then the $k = 0$ Friedmann model. This is the Friedmann limit of the metric (2.1), as expected for Bianchi type III.

In this case, the coordinates of the Friedmann limit are similar to those usually used (they are the nonorthogonal Cartesian coordinates for the flat space $t = $ const). This will not be so in most other cases – the coordinate representation of the resulting Friedmann limit will be rather exotic, and calculating the Riemann tensor of the subspace $t = $ const will be the simplest way to check that it is the Friedmann metric indeed.

The Killing fields for the metric (2.1) are (see Ref. 2):

$$k^\alpha_{(1)} = \delta^\alpha_1, \qquad k^\alpha_{(2)} = e^{bx}(\delta^\alpha_0 - b\delta^\alpha_2), \qquad k^\alpha_{(3)} = C_3\delta^\alpha_0 + \lambda_3\delta^\alpha_3. \tag{2.22}$$

As seen from Fig. 1, the algebra of type III can be specialized only to types II and I, and so the $k = 0$ Friedmann limit is the only one of the three that can be expected here. Note that the Killing field $k_{(2)}$ will have a meaninigful limit $\omega \to 0$, $b \to 0$ only if the two limits are tuned so that $\omega_0/b \xrightarrow[\omega_0 \to 0]{} 0$ (for example, $b = B\sqrt{\omega_0}$). Then $l^\alpha_{(2)} := (\omega_0/b)k^\alpha_{(2)} \xrightarrow[\omega_0 \to 0]{} \delta^\alpha_2$, which is indeed a Killing filed of (2.18) with $b = 0$. The algebra $\{k_{(1)}, l_{(2)}, k_{(3)}\}$ becomes then Bianchi type I when $\omega_0 = 0$, as expected.

The reasoning behind the reparametrizations, and the subsequent calculation of the limits of zero rotation, zero tilt, zero shear and constant curvature of the spaces $t = $ const, follows the same scheme in all the other cases. Therefore, it will be presented in less detail from now on. In some of the cases, the reparametrization that untilts the limit $\omega \to 0$ is a necessary condition for cancelling the infinities introduced by the earlier reparametrizations.

The other two subcases of case 1.1.1 in Ref. 2, i.e. cases 1.1.1.2.1 (eq. (3.16)) and 1.1.1.2.2 (eq. (3.32)) are immediately seen to allow no Friedmann limit. For both of them, the Killing fields are given by (2.22) above with $\lambda_3 = 0$. As seen from (1.8), the Killing field $k_{(3)}$ coincides then with the velocity field of dust, and so both these models are stationary.

In fact, the last of (2.22) is the linear relation among the five vectors mentioned at the end of sec. I, becuase it is equivalent to (1.12). Since with $\lambda_3 = 0$ the velocity becomes one of the Killing fields, i.e. becomes tangent to the symmetry orbits, $\lambda_3$ is the tilt parameter. More on this – see sec. V.

# III. Cases 1.1.2 of Ref. 2.

The case 1.1.2.1 is again of Bianchi type III. The transformation back from the coordinates of eqs. (4.6) in Ref. 2 (adapted to two Killing fields) to the Plebański coordinates is given by (4.4) in Ref. 2, with the roles of $\{x^\alpha\}$ and $\{x'^\alpha\}$ interchanged. The transformed metric is:

$$g_{11} = -2(c/a)y - (c/a)^2 + Y^2 h_{11} - 2(c/a)\lambda_3 Y h_{13} + (c\lambda_3/a)^2 h_{33},$$

$$g_{12} = h_{12} - \frac{c\lambda_3}{aY} h_{23}, \qquad g_{13} = C_3 Y h_{13} - (c/a)\lambda_3 C_3 h_{33},$$

$$g_{22} = h_{22}/Y^2, \qquad g_{23} = C_3 h_{23}/Y, \qquad g_{33} = C_3{}^2 h_{33}, \qquad Y := ay + c, \qquad (3.1)$$

where $a$, $c$ and $\lambda_3$ are arbitrary constants and $h_{ij}$ are arbitrary functions of the variable:

$$T := t - C_3 z/\lambda_3. \qquad (3.2)$$

The reparametrization that will allow setting the rotation and tilt to zero is:

$$(y, C_3) = \omega_0(\tilde{y}, D),$$

$$h_{11} = H_{11} + (\lambda_3/a)^2 H_{33}/\omega_0{}^2, \qquad h_{12} = H_{12}/\omega_0 + (\lambda_3/a)H_{23}/\omega_0{}^2,$$

$$h_{13} = H_{13} + (\lambda_3/a)H_{33}/\omega_0{}^2, \qquad (h_{22}, h_{23}, h_{33}) = (H_{22}, H_{23}, H_{33})/\omega_0{}^2. \qquad (3.3)$$

The reparametrized metric (without the limit $\omega_0 \to 0$ taken yet) is:

$$g_{11} = -(c/a)^2 - 2(c/a)\omega_0 \tilde{y} + \tilde{Y}^2 H_{11} - 2(c/a)\lambda_3 \tilde{Y} H_{13} + (\lambda_3 \tilde{y})^2 H_{33}$$

$$g_{12} = H_{12} + (\lambda_3 \tilde{y}/\tilde{Y})H_{23}, \qquad g_{13} = D\omega_0 \tilde{Y} H_{13} + D\lambda_3 \tilde{y} H_{33},$$

$$g_{22} = H_{22}/\tilde{Y}^2, \qquad g_{23} = DH_{23}/\tilde{Y}, \qquad g_{33} = D^2 H_{33}, \qquad \tilde{Y} = a\omega_0 \tilde{y} + c. \qquad (3.4)$$

Similarly as before, in the limit $\omega_0 \to 0$ the $H_{ij}$ will depend only on $t$, and the subsequent limit of zero shear is $H_{11} = -C_{11}R^2(t) + (c/a)^2$, other $H_{ij}(t) = -C_{ij}R^2(t)$, $C_{33} = 1$. Proceeding exactly as in sec. II, we then find that the hypersurfaces $t = \text{const}$ will have constant curvature when $\lambda_3 \to 0$; the resulting limit is the Friedmann $k = 0$ model, as expected for type III. The limits $C_3 \to 0$ and $\lambda_3 \to 0$ should be tuned so that $C_3/\lambda_3 \xrightarrow[\omega_0 \to 0]{} 0$, e.g. $\lambda_3 = L_3 \omega_0{}^{1/2}$.

The case 1.1.2.2 (eqs. (4.12) – (4.33) in Ref. 2) is of Bianchi type II. It is known from the paper by Ozsváth[7], and from Theorem 3.1 by King and Ellis[8], that dust models of type II have zero rotation. However, that thesis was proven with use of the Einstein equations in Ref. 7 and of the Ellis evolution equations[9] in Ref. 8, that include consequences of the Einstein equations. In the approach of Refs. 1 – 3, the Einstein equations were not used. Moreover, the constant $\lambda_3$ plays the role of the tilt parameter here – with $\lambda_3 = 0$, the metric becomes stationary (the orbits of the symmetry group become timelike and tangent to the velocity field of the dust), and this case is not covered in Refs. 7 and 8. This is why the case 1.1.2.2 could show up in our consideration. This observation implies a warning: the existence of a Friedmann limit of the metric does not guarantee that the Einstein equations will allow a rotating generalization of a given Bianchi type of the corresponding Friedmann model. A rotating dust solution and the Friedmann solution may turn out to be two disjoint subclasses within that type.

The metric (eq. (4.13) from Ref. 2) transformed back to the Plebański coordinates (by the inverse of (4.4) from Ref. 2) is:

$$g_{11} = h_{11} + 2\lambda_3 y h_{13} + y^2(1 + \lambda_3{}^2 h_{33})$$

$$g_{12} = h_{12} + \lambda_3 y h_{23}, \qquad g_{13} = C_3(h_{13} + \lambda_3 y h_{33}),$$

$$(g_{22}, g_{23}, g_{33}) = (h_{22}, C_3 h_{23}, C_3{}^2 h_{33}), \tag{3.5}$$

where the $h_{ij}$ are arbitrary functions of the $T$ from (3.2). The limit of zero rotation and zero tilt is achieved after the reparametrization:

$$(y, C_3) = \omega_0(\tilde{y}, D),$$

$$(h_{12}, h_{13}) = (H_{12}, H_{13})/\omega_0, \qquad (h_{22}, h_{23}, h_{33}) = (H_{22}, H_{23}, H_{33})/\omega_0{}^2. \tag{3.6}$$

and the reparametrized metric is:

$$g_{11} = (\omega_0 \tilde{y})^2 + h_{11} + 2\lambda_3 \tilde{y} H_{13} + (\lambda_3 \tilde{y})^2 H_{33}$$

$$g_{12} = H_{12} + \lambda_3 \tilde{y} H_{23}, \qquad g_{13} = D(H_{13} + \lambda_3 \tilde{y} H_{33}),$$

$$(g_{22}, g_{23}, g_{33}) = (H_{22}, DH_{23}, D^2 H_{33}). \tag{3.7}$$

The $k = 0$ Friedmann limit will result now when $\omega_0 = 0$, $H_{ij} = -C_{ij} R^2$ and $\lambda_3 = 0$.

The theorem by King and Ellis mentioned above implies that $\omega_0 = 0$ will follow when (3.7) is substituted in the Einstein equations.

The Killing fields for the metric (3.1) are:

$$k^\alpha_{(1)} = \delta^\alpha{}_1, \qquad k^\alpha_{(3)} = C_3 \delta^\alpha{}_0 + \lambda_3 \delta^\alpha{}_3,$$

$$k^\alpha_{(2)} = cx\delta^\alpha{}_0 + ax\delta^\alpha{}_1 - (ay + c)\delta^\alpha{}_2 + (c\lambda_3/C_3)x\delta^\alpha{}_3. \tag{3.8}$$

(The Killing fields for (3.5) result when $a = 0$ above.) After the reparametrization (3.3), in the limit $\omega_0 \to 0$, the basis (3.8) becomes:

$$k^\alpha_{(1)} = \delta^\alpha{}_1, \qquad l^\alpha_{(3)} := (1/\lambda_3) k^\alpha_{(3)} \xrightarrow[\omega_0 \to 0]{} \delta^\alpha{}_3$$

$$l^\alpha_{(2)} = -(\omega_0/c) k^\alpha_{(2)} \xrightarrow[\omega_0 \to 0]{} \delta^\alpha{}_2 - (\lambda_3/D) x \delta^\alpha{}_3. \tag{3.9}$$

In the Friedmann limit $\lambda_3 \to 0$, the generators (3.9) become a Bianchi type I algebra.

## IV. Cases 1.2 and 2 of Ref. 2.

All of these allow both the $k = 0$ and the $k = -1$ Friedmann limits.

Case 1.2.1.1 is of Bianchi type $VI_h$ with the free parameter $(b^2 + f^2)/(b^2 - f^2)$ (there is a typo in Ref. 2). In this case (eqs. (5.6) – (5.7) in Ref. 2), the transformation back to the Plebański coordinates is given by (5.5) from Ref. 2, with the roles of $x^\alpha$ and $x'^\alpha$ interchanged. The resulting metric is:

$$g_{11} = h_{11} - 2bt(y - bh_{12}) - 2Zh_{13} + (bt)^2(b^2 h_{22} - 1) - 2b^2 t Z h_{23} + Z^2 h_{33},$$

$$g_{12} = h_{12} + b^2 t h_{22} - Z h_{23}, \qquad g_{13} = h_{13} + b^2 t h_{23} - Z h_{33},$$

$$(g_{22}, g_{23}, g_{33}) = (h_{22}, h_{23}, h_{33}), \qquad Z := fz, \qquad (4.1)$$

where $b$ and $f$ are arbitrary constants, and $h_{ij}$ are arbitrary functions of:

$$T = t + y/b. \qquad (4.2)$$

The limit of zero rotation (that will automatically untilt the model) is achieved by the reparametrization:

$$y = \omega_0 \tilde{y}, \qquad h_{22} = H_{22}/\omega_0^2, \qquad h_{23} = H_{23}/\omega_0,$$

$$h_{11} = H_{11} + (bT)^2 - 2b^2 T h_{12} - b^4 T^2 H_{22}/\omega_0^2,$$

$$h_{12} = H_{12}/\omega_0 - b^2 T H_{22}/\omega_0^2, \qquad h_{13} = H_{13} - b^2 T H_{23}/\omega_0, \qquad (4.3)$$

and the reparametrized metric is:

$$g_{11} = (\omega_0 \tilde{y})^2 + H_{11} - 2b\tilde{y} H_{12} - 2Z H_{13} + (b\tilde{y})^2 H_{22} + 2b\tilde{y} Z H_{23} + Z^2 h_{33},$$

$$g_{12} = H_{12} - b\tilde{y} H_{22} - Z H_{23}, \qquad g_{13} = H_{13} - b\tilde{y} H_{23} - Z h_{33},$$

$$(g_{22}, g_{23}, g_{33}) = (H_{22}, H_{23}, h_{33}). \qquad (4.4)$$

With $\omega_0 = 0$, the shearfree limit will result when all $h_{ij} = -C_{ij} R^2(t)$, and then the $k = -1$ Friedmann model results when $b = f \neq 0$. The $k = 0$ Friedmann limit results when $b = f = 0$. This is the first instance where the coordinates of the $k = -1$ Friedmann limit come out rather exotic. From now on, this will be the rule.

The Killing fields for the metric (4.1) – (4.2) are:

$$k^\alpha_{(1)} = \delta^\alpha{}_1, \qquad k^\alpha_{(2)} = \mathrm{e}^{bx}(\delta^\alpha{}_0 - b\delta^\alpha{}_2), \qquad k^\alpha_{(3)} = \mathrm{e}^{fx}\delta^\alpha{}_3. \qquad (4.5)$$

In the $k = -1$ Friedmann limit that will result by the first of (4.3) and $b = f$, $k^\alpha_{(1)}$ remains unchanged, $k^\alpha_{(3)}$ becomes $\mathrm{e}^{bx}\delta^\alpha{}_{(3)}$, while $k^\alpha_{(2)}$ is replaced by:

$$l^\alpha_{(2)} = (-\omega_0/b) k^\alpha_{(2)} \xrightarrow[\omega_0 \to 0]{} \mathrm{e}^{bx}\delta^\alpha{}_2. \qquad (4.6)$$

This is of Bianchi type V, and in the further limit $b = f = 0$ it becomes type I.

In the case 1.2.1.2 (eqs. (5.10) in Ref. 2), which is of type IV, the transformation back to the Plebański coordinates is given by eq. (5.9) there. The whole further calculation is similar to (4.1) – (4.4) above. Instead of the last formula in (4.1) we have:

$$Z := ct + bz, \qquad (4.7)$$

where $c$ is one more arbitrary constant, and in (4.3) we have:

$$h_{11} = H_{11} + (bT)^2 - 2b^2 T h_{12} + 2cT H_{13} - b^4 T^2 H_{22}/\omega_0^2 + (cT)^2 h_{33},$$

$$h_{12} = H_{12}/\omega_0 - b^2 T H_{22}/\omega_0^2 + cT H_{23}/\omega_0, \qquad h_{13} = H_{13} - b^2 T H_{23}/\omega_0 + cT h_{33}. \qquad (4.8)$$

The reparametrized metric is:

$$g_{11} = (\omega_0 \tilde{y})^2 (1 + c^2 h_{33}/b^2) + 2c\omega_0 \tilde{y}(H_{13}/b - \tilde{y} H_{23} - z h_{33})$$

$$+H_{11} - 2b\tilde{y}H_{12} - 2bzH_{13} + (b\tilde{y})^2 H_{22} + 2b^2\tilde{y}zH_{23} + (bz)^2 h_{33},$$

$$g_{12} = \omega_0(c/b)\tilde{y}H_{23} + H_{12} - b\tilde{y}H_{22} - bzH_{23},$$

$$g_{13} = \omega_0(c/b)\tilde{y}h_{33} + H_{13} - b\tilde{y}H_{23} - bzh_{33}, \qquad (g_{22}, g_{23}, g_{33}) = (H_{22}, H_{23}, H_{33}). \qquad (4.9)$$

The limit $\omega_0 \to 0$ of (4.9) is the same as the limit $\omega_0 \to 0$ of (4.4) with $b = f$. Hence, the $k = -1$ Friedmann limit will result from (4.9) when $\omega_0 = 0$ and $h_{ij} = -C_{ij}R^2(t)$, without any further limitations. The $k = 0$ Friedmann limit will result when $b = 0$ in addition.

For the case 1.2.2.1 (eqs. (5.17) – (5.18) in Ref. 2), which is of Bianchi type $VI_h$, the subcase $C = j + a = 0$ is identical to the subcase $c = 0$ of case 1.2.2.2, and so the Friedmann limits will be the same (see below).

The case 1.2.2.2 (eq. (5.19) in Ref. 2), which is of Bianchi type IV, allows the special case $c = 0$, where the Bianchi type becomes V. This special case was investigated in detail in Ref. 4, and it was shown there how the $k = -1$ Friedmann limit is obtained. In order to obtain the $k = 0$ Friedmann limit, one has to apply the following transformation and rescaling to eq. (3.5) in Ref. 4:

$$y = e^{\alpha u}, \qquad K = \tilde{K}/\alpha, \qquad (4.10)$$

and then take the limit $\alpha \to 0$.

All the subcases of case 2 in Ref. 2 have matter density constant along the dust flow: in case 2.1.1 (type I) and both cases 2.1.2 (types II and I), the velocity field is tangent to the symmetry orbits, in case 2.2 (type I), the velocity field coincides with one of the Killing fields. Therefore, no Friedmann limits will exist there.

With this, all cases of Ref. 2 are exhausted.

## V. Case 1.1.1.1 of Ref. 3.

In the cases considered in Ref. 3, each of the 3 Killing vectors is linearly independent of the velocity and rotation. However, the 5 vectors existing in each 4-dimensional tangent space to the manifold cannot form a linearly independent set. The 3-dimensional space spanned by the Killing vectors, $K_3$, must intersect with the 2-dimensional plane spanned by the velocity and rotation, $H_2$, along at least one direction. In the models of Ref. 1, the whole $H_2$ plane was a subspace of the $K_3$ space. In consequence, the velocity was a linear combination of the Killing vectors, and so those models were stationary. In the models of Ref. 2, considered up to now, the plane $H_2$ and the space $K_3$ intersected along the direction of the Killing vector $k_{(3)}^\alpha = C_3 u^\alpha + (\lambda_3/n)w^\alpha$. From now on, the line of intersection will not coincide with any Killing direction. Hence, in each case an equation of the following form will have to hold:

$$a_1 k_{(1)}^\alpha + a_2 k_{(2)}^\alpha + a_3 k_{(3)}^\alpha = b_1 u^\alpha + b_2 w^\alpha, \qquad (5.1)$$

where $a_i$ and $b_i$ are functions on the manifold. Note that if $b_2 = 0$, then the velocity field is tangent to the symmetry orbits, and in consequence such a model has zero expansion and matter-density independent of the comoving time (the metric may depend on the time only because in general the metric has shear). Hence, $b_2$ is a measure of the tilt of the velocity field with respect to the orbits. Its relation to the tilt defined by King and Ellis[8] will be explained below (see after eq. (5.6)).

The case 1.1.1.1, given by eqs. (2.28) – (2.29) in Ref. 3, is of Bianchi type $VI_h$. The transformation back to the Plebański coordinates is given by eq. (2.27) in Ref. 3, and the result is:

$$g_{11} = \frac{f^2(b+f)}{b^2(b-f)}U^2 - 2\frac{f}{b-f}U(ft+y) + h_{11} - 2Vh_{12} - 2\frac{f\gamma}{b\beta(b-f)}Uh_{13}$$

$$+V^2 h_{22} + 2\frac{f\gamma}{b\beta(b-f)}UVh_{23} + \left[\frac{f\gamma}{b\beta(b-f)}\right]^2 U^2 h_{33},$$

$$g_{12} = V/b^2 + h_{12} - Vh_{22} - \frac{f\gamma}{b\beta(b-f)}Uh_{23},$$

$$g_{13} = h_{13} - Vh_{23} - \frac{f\gamma}{b\beta(b-f)}Uh_{33},$$

$$g_{22} = -1/b^2 + h_{22}, \qquad (g_{23}, g_{33}) = (h_{23}, h_{33}), \tag{5.2}$$

where $b$, $f$, $\beta$ and $\gamma$ are arbitrary constants, the $h_{ij}$ are arbitrary functions of the argument:

$$T = t + y/b - \beta(b-f)z/\gamma, \tag{5.3}$$

and $U$ and $V$ stand for:

$$U = bt + y, \qquad V = bft + (b+f)y. \tag{5.4}$$

The Killing fields for the metric (5.2) are:

$$k^\alpha_{(1)} = \delta^\alpha_1, \qquad k^\alpha_{(2)} = e^{fx}\left\{\delta^\alpha_0 - f\delta^\alpha_2 + [\gamma/(b\beta)]\delta^\alpha_3\right\},$$

$$k^\alpha_{(3)} = e^{bx}(-\delta^\alpha_0 + b\delta^\alpha_2). \tag{5.5}$$

From (1.8) it follows then that:

$$be^{-fx}k^\alpha_{(2)} + fe^{-bx}k^\alpha_{(3)} = (b-f)u^\alpha + [\gamma/(n\beta)]w^\alpha. \tag{5.6}$$

This is the equation (5.1) specified for the case 1.1.1.1. As remarked above, when $\gamma = 0$, the velocity field becomes tangent to the symmetry orbits. (With $\gamma \to 0$, the argument of $h_{ij}$ given by (5.2) has to be redefined so that it becomes $Z = \gamma T \xrightarrow[\gamma \to 0]{} -\beta(b-f)z$.)

This means that the parameter $(\gamma/\beta)$ is a measure of the tilt of the velocity field with respect to the symmetry orbits. Indeed, the measure of tilt defined by King and Ellis[8] is proportional to $(\gamma/\beta)$. They defined the hyperbolic angle of tilt $\overline{\beta}$ by:

$$\cosh\overline{\beta} = u^\alpha n_\alpha \tag{5.7}$$

(the difference in sign from their paper is a consequence of the difference in signature), where $n_\alpha$ is the unit vector normal to the orbits of symmetry. This definition of $\overline{\beta}$ makes sense only when both $u^\alpha$ and $n^\alpha$ are timelike vectors; the cases of $n^\alpha$ being null or spacelike are not considered in Ref. 8. However, $u^\alpha n_\alpha$ is a measure of the tilt also for nontimelike $n^\alpha$. In particular, when $u^\alpha$ is tangent to the symmetry orbits, $u^\alpha n_\alpha = 0$. The vector $n^\alpha$ is related to the Killing fields by:

$$n_\alpha = N_\alpha / \sqrt{-g_{\mu\nu} N^\mu N^\nu} := N_\alpha / \|N\|, \tag{5.8}$$

where:

$$N_\alpha = \frac{1}{\sqrt{-g}} \varepsilon_{\alpha\beta\gamma\delta} k^\beta_{(1)} k^\gamma_{(2)} k^\delta_{(3)}. \tag{5.9}$$

In our case then:

$$u^\alpha n_\alpha = \|N\|^{-1} N_\alpha u^\alpha = \frac{1}{\|N\|\sqrt{-g}} e^{(b+f)x} \gamma/\beta. \tag{5.10}$$

Analogs of (5.1) and (5.6) will exist in every case considered from now on. In the models of Ref. 2, considered up to now, where the Killing field $k^\alpha_{(3)}$ always had the form $k^\alpha_{(3)} = C_3 u^\alpha + (\lambda_3/n) w^\alpha$, $\lambda_3$ was a measure of the tilt.

For calculating the limit of zero rotation and zero shear, the following reparametrization is useful:

$$(y, \beta) = \omega_0(\tilde{y}, B), \qquad h_{22} = H_{22}/\omega_0^2, \qquad h_{23} = H_{23}/\omega_0,$$
$$h_{11} = (fT)^2 + H_{11} + 2bfTh_{12} + 2(f^2/D)Th_{13}/\omega_0$$
$$- (bfT)^2 H_{22}/\omega_0^2 - 2(bf^3/D)T^2 H_{23}/\omega_0^2 - (f^4/D^2)T^2 h_{33}/\omega_0^2,$$
$$h_{12} = H_{12}/\omega_0 + bfTH_{22}/\omega_0^2 + (f^2/D)TH_{23}/\omega_0^2,$$
$$h_{13} = H_{13} + bfTH_{23}/\omega_0 + (f^2/D)Th_{33}/\omega_0,$$
$$D := f(b-f)B/\gamma. \tag{5.11}$$

The reparametrized metric is:

$$g_{11} = -2ftW\omega_0 + [(Dz)^2 - 2(f/b)\tilde{y}W]\omega_0^2 + H_{11} - 2bWH_{12}$$
$$-2fzH_{13} + (bW)^2 H_{22} + 2bfzWH_{23} + (fz)^2 H_{33},$$
$$g_{12} = (ft/b)\omega_0 + [(b+f)/b^2]\tilde{y}\omega_0^2 + H_{12} - bWH_{22} - fzH_{23},$$
$$g_{13} = H_{13} - bWH_{23} - fzh_{33},$$
$$(g_{22}, g_{23}, g_{33}) = (H_{22}, H_{23}, h_{33} = H_{33}), \qquad W := y + Dz. \tag{5.12}$$

The $k = -1$ Friedmann limit results now from (5.12) when $\omega_0 = 0$ (after which all $h_{ij}$ depend only on $t$), and:

$$H_{ij} = -C_{ij} R^2(t), \qquad C_{33} = 1, \qquad b = f. \tag{5.13}$$

(The last of (5.13) implies $D = 0$.) The $k = 0$ Friedmann limit results when in addition:

$$b = f = 0. \tag{5.14}$$

The reparametrization (5.11) transforms the Killing fields $k^\alpha_{(2)}$ and $k^\alpha_{(3)}$ from (5.5) as follows:

$$l^\alpha_{(2)} = (b\beta/\gamma) k^\alpha_{(2)} \xrightarrow[\omega_0 \to 0]{} e^{fx}[-(bf\beta/\gamma)\delta^\alpha_{(2)} + \delta^\alpha_{(3)}]$$
$$l^\alpha_{(3)} = (\omega_0/b) k^\alpha_{(3)} \xrightarrow[\omega_0 \to 0]{} e^{bx} \delta^\alpha_{(2)}. \tag{5.15}$$

In the $k = -1$ Friedmann limit ($b = f$), together with $k^\alpha_{(1)}$, this becomes a Bianchi type V algebra, and in the $k = 0$ Friedmann limit ($b = f = 0$), it becomes type I.

# VI. Case 1.1.1.2: of Ref. 3.

This case is given by eqs. (3.7) – (3.11) in Ref. 3. It is of Bianchi type VIII or $VI_0$ (when $g \neq 0$ or $g = 0$, respectively), so only the $k = 0$ Friedmann limit may exist here. The limit of zero rotation and zero tilt can be considered without transforming the metric back to the Plebański form, but the 3 subcases have to be considered separately.

The Killing fields in this case are:

$$k^{\alpha}_{(1)} = \delta^{\alpha}_1, \qquad k^{\alpha}_{(3)} = e^{\alpha_1 x}\delta^{\alpha}_2,$$

$$k^{\alpha}_{(2)} = e^{-\alpha_1 x}[2gy\delta^{\alpha}_1 + \alpha_1(gy^2 + 2B)\delta^{\alpha}_2 + \delta^{\alpha}_3]. \tag{6.1}$$

The analog of eq. (5.1) is:

$$k^{\alpha}_{(2)} - \alpha_1(gy^2 + 2B)e^{-2\alpha_1 x}k^{\alpha}_{(3)} - 2gye^{-\alpha_1 x}k^{\alpha}_{(1)} = e^{-\alpha_1 x}[4Bu^{\alpha} + (8c\gamma\alpha_1/n)w^{\alpha}], \tag{6.2}$$

and, consequently, the King–Ellis measure of tilt is:

$$\sqrt{-g}N_{\alpha}u^{\alpha} = \alpha_1 \tag{6.3}$$

($N_{\alpha}$ is given by (5.8)).

The argument of the arbitrary functions in the metric is:

$$T = t + y/\alpha_1 - \frac{B}{2c\gamma\alpha_1}z. \tag{6.4}$$

In case I ($gB \neq 0$), the reparametrization needed is:

$$y = \omega_0\tilde{y}, \qquad B = \omega_0^{3/4}\tilde{B}, \qquad \alpha_1 = \omega_0^{1/4}a_1, \tag{6.5}$$

$$(h_{13}, k_{13}) = (G_{13}, K_{13})\omega_0 \qquad h_{22} = G_{22}/\omega_0^2, \tag{6.6}$$

The full reparametrized metric, with $\omega_0 \neq 0$, is rather complicated here, so only the limit $\omega_0 \to 0$ will be quoted:

$$ds^2 = dt^2 + h_{11}dx^2 + G_{22}dy^2 - \frac{h_{33}}{2\tilde{B}a_1}dydz + h_{33}dz^2. \tag{6.7}$$

The $k = 0$ Friedmann limit results when further $h_{11} = G_{22} = h_{33} = -R^2(t)$.

As seen from (6.1), the symmetry group becomes Bianchi type I in the limit $\omega_0 \to 0$ after the reparametrization (6.4) (the Killing field $k_{(3)}$ has to be replaced by $l_{(3)} = \omega_0 k_{(3)}$ in order that the limit is nonsingular).

In case II ($B = 0$), eq. (6.5) remains unchanged, while (6.6) is replaced by:

$$(h_{12}, h_{23}) = (G_{12}, G_{23})\omega_0 \qquad h_{22} = G_{22}/\omega_0^2. \tag{6.8}$$

The limit $\omega_0 \to 0$ of the reparametrized metric is:

$$ds^2 = dt^2 + h_{11}dx^2 + 2G_{12}dxdy + 2h_{13}dxdz + G_{22}dy^2 + h_{33}dz^2, \tag{6.9}$$

where all the metric components depend only on $t$. The $k = 0$ Friedmann limit is here

$$G_{12} = h_{13} = 0, \qquad h_{11} = G_{22} = h_{33} = -R^2(t). \tag{6.10}$$

In case III ($g = 0$, Bianchi type VI$_0$), the $k = 0$ Friedmann limit results again by (6.5), (6.8), (6.9) and (6.10).

Note two typos in Sec. 3 of Ref. 3: in (3.7) the correct formula for $w^\alpha$ is:

$$w^\alpha = \frac{n\alpha_1}{2c\gamma\Delta}(-4B\delta_0^\alpha + \delta_3^\alpha) = \frac{n}{8c\gamma\alpha_1}(-4B\delta_0^\alpha + \delta_3^\alpha), \tag{6.11}$$

and in (3.10), the correct formula for $g_{23}$ is:

$$g_{23} = -2gzh_{13} + h_{23} - \alpha_1 g z^2 h_{33}. \tag{6.12}$$

## VII. Case 1.1.2.1 of Ref. 3.

This model is of Bianchi type VII$_h$, and is given by eqs. (4.19) – (4.23) in Ref. 3. Two formulae in (4.23) had typos, the correct expressions are:

$$g_{12} = e^{(b+f)x/2}[Wh_{12} - (\gamma/D)(b+f)\cos(Dx/2)h_{13}],$$

$$g_{22} = e^{(b+f)x}\{[\gamma^2(b+f)^2 h_{33}/D^2 + 1]\cos^2(Dx/2)$$
$$-2(\gamma/D)(b+f)\cos(Dx/2)Wh_{23} + W^2 h_{22}\} \tag{7.1}$$

The transformation back to the Plebański coordinates is given by (4.21) in Ref. 3, and the resulting metric is:

$$g_{11} = y^2 + h_{11} + 2Uh_{12} - 2(\gamma/D)(b+f)yh_{13} + U^2 h_{22}$$
$$-2(\gamma/D)(b+f)yUh_{23} + [(\gamma/D)(b+f)y]^2 h_{33},$$
$$g_{12} = -2h_{12} + 2(\gamma/D)h_{13} - 2Uh_{22} + 2(\gamma/D)Uh_{23}$$
$$+2(\gamma/D)(b+f)yh_{23} - 2(\gamma/D)^2(b+f)yh_{33},$$
$$g_{13} = h_{13} + Uh_{23} - (\gamma/D)(b+f)yh_{33},$$
$$g_{22} = 4h_{22} - 8(\gamma/D)h_{23} + 4(\gamma/D)^2 h_{33}, \qquad g_{23} = -2h_{23} + 2(\gamma/D)h_{33},$$
$$g_{33} = h_{33}, \qquad U := \frac{1}{2}[(b+f)^2 + D^2]t + 2(b+f)y, \tag{7.2}$$

where $b$, $f$ $D$ and $\gamma$ are arbitrary constants, and $h_{ij}$ are arbitrary functions of the variable:

$$T = t + \frac{2y}{b+f} + \frac{D}{\gamma(b+f)}z. \tag{7.3}$$

The Killing fields for the metric (7.2) are:

$$k_{(1)}^\alpha = \delta_1^\alpha, \qquad k_{(2)}^\alpha = e^{(b+f)x/2}[\cos(Dx/2)\delta_0^\alpha - \frac{1}{2}W\delta_2^\alpha - \gamma\sin(Dx/2)\delta_3^\alpha],$$

$$k^\alpha_{(3)} = e^{(b+f)x/2}[\sin(Dx/2)\delta^\alpha_0 - \frac{1}{2}V\delta^\alpha_2 + \gamma\cos(Dx/2)\delta^\alpha_3],$$

$$W := (b+f)\cos(Dx/2) - D\sin(Dx/2),$$

$$V := D\cos(Dx/2) + (b+f)\sin(Dx/2). \tag{7.4}$$

The analog of (5.1) is here:

$$Vk^\alpha_{(2)} - Wk^\alpha_{(3)} = De^{(b+f)x/2}u^\alpha - \frac{\gamma}{n}(b+f)e^{(b+f)x/2}w^\alpha, \tag{7.5}$$

and the King–Ellis measure of the tilt is:

$$\sqrt{-g}u^\alpha N_\alpha = -\frac{1}{2}(b+f)\gamma e^{(b+f)x}. \tag{7.6}$$

Eq. (7.5) shows that with $\gamma(b+f) = 0$, the model should be nonexpanding. This is so indeed, but in order to be able to consider the subcase $\gamma(b+f) \to 0$, we have to take $\gamma(b+f)T$ as the argument of $h_{ij}$ in (7.2) instead of the $T$ given by (7.3).

We define:
$$E := (b+f)^2 + D^2, \tag{7.7}$$

and then the reparametrization needed for the limit of zero rotation and zero tilt is:

$$(y, D) = \omega_0(\tilde{y}, d), \qquad h_{22} = H_{22}/\omega_0^2, \qquad h_{23} = H_{23}/\omega_0,$$

$$h_{11} = H_{11} - ETH_{12}/\omega_0 + \left(\frac{1}{2}ET\right)^2 H_{22}/\omega_0^2,$$

$$h_{12} = H_{12}/\omega_0 - \frac{1}{2}ETH_{22}/\omega_0^2, \qquad h_{13} = H_{13} - \frac{1}{2}ETH_{23}/\omega_0. \tag{7.8}$$

After the reparametrization we have:

$$S_{1,2} := b + f + \varepsilon_{1,2}(\omega_0 d)^2/(b+f) \xrightarrow[\omega_0 \to 0]{} b+f, \qquad \varepsilon_1 = +1, \varepsilon_2 = -1,$$

$$\Sigma := S_2\tilde{y} - \frac{1}{2}(d/\gamma)S_1 z,$$

$$g_{11} = (\omega_0 y)^2 + H_{11} + 2\Sigma H_{12} - 2(\gamma/d)(b+f)\tilde{y}H_{13} + \Sigma^2 H_{22}$$
$$-2(\gamma/d)(b+f)\tilde{y}\Sigma H_{23} + [(\gamma/d)(b+f)\tilde{y}]^2 h_{33},$$

$$g_{12} = -2H_{12} + 2(\gamma/d)H_{13} - 2\Sigma H_{22} + 2(\gamma/d)\tilde{y}H_{23}$$
$$+2(\gamma/d)(b+f+\Sigma)H_{23} - 2(\gamma/d)^2(b+f)\tilde{y}h_{33},$$

$$g_{13} = H_{13} + \tilde{y}\Sigma H_{23} - (\gamma/d)(b+f)\tilde{y}h_{33},$$

$$g_{22} = 4H_{22} - 8(\gamma/d)H_{23} + 4(\gamma/d)^2 h_{33},$$

$$g_{23} = -2H_{23} + 2(\gamma/d)h_{33}, \qquad g_{33} = h_{33} := H_{33}. \tag{7.9}$$

In the limit $\omega_0 \to 0$, all the $H_{ij}$ will depend only on $t$. The limit of zero shear is then obtained by:

$$H_{ij} = -C_{ij}R^2(t), \qquad C_{33} = 1. \tag{7.10}$$

To obtain the Friedmann limits, a further reparametrization of the constants $C_{ij}$ is necessary. We define:
$$C_{23} = D_{23} + \gamma/d, \qquad D_{22}{}^2 = C_{22} - C_{23}{}^2,$$
$$D_{12} = (C_{12} - C_{13}C_{23})/D_{22}, \qquad D_{11}{}^2 = C_{11} - C_{13}{}^2 - D_{12}{}^2. \tag{7.11}$$

The metric (7.9) may then be written:
$$\mathrm{d}s^2 = \mathrm{d}t^2 - (D_{11}R\mathrm{d}x)^2 - R^2\left\{\left[-D_{12} - (b+f)D_{22}\left(\tilde{y} - \frac{d}{2\gamma}z\right)\right]\mathrm{d}x + 2D_{22}\mathrm{d}\tilde{y}\right\}^2$$
$$-R^2\left\{\left[C_{13} + (b+f)D_{23}\tilde{y} - \frac{1}{2}(b+f)\left(1 + \frac{d}{\gamma}D_{23}\right)z\right]\mathrm{d}x - 2D_{23}\mathrm{d}\tilde{y} + \mathrm{d}z\right\}^2, \tag{7.12}$$

The $k = -1$ Friedmann limit results now when $d = 0$, the $k = 0$ limit results when $b+f = 0$ in addition. The first of (7.11) was necessary to eliminate $\gamma/d$ from (7.9) so that the limit $d \to 0$ could be subsequently taken.

We have found above (after eq. (7.6)) that $b + f = 0$ corresponds to zero expansion. This is so when $b + f \to 0$ with other parameters uchanged. In considering the $k = 0$ Friedmann limit, $b+f$ is set to zero after the limit $d \to 0$ had already been taken. In order to make these two limits compatible, we have to assume that $b + f \to 0$ slowly enough so that $D/(b+f) \to 0$ and $y/(b+f) \to 0$. With the reparametrization (7.8), this is achieved when $b + f = B\omega_0{}^\varepsilon$, where $0 < \varepsilon < 1$.

After the reparametrization (7.8), in the limit $\omega_0 \to 0$, the Killing fields become:
$$k^\alpha_{(1)} = \delta^\alpha{}_{(1)}, \qquad l^\alpha_{(2)} = -[2\omega_0/(b+f)]k^\alpha_{(2)} \xrightarrow[\omega_0 \to 0]{} \mathrm{e}^{(b+f)x/2}\delta^\alpha{}_{(2)}$$
$$l^\alpha_{(3)} = (1/\gamma)k^\alpha_{(3)} \xrightarrow[\omega_0 \to 0]{} \mathrm{e}^{(b+f)x/2}\left\{-\frac{d}{2\gamma}\left[1 + \frac{1}{2}(b+f)x\right]\delta^\alpha{}_{(2)} + \delta^\alpha{}_{(3)}\right\}. \tag{7.13}$$

This is a Bianchi type IV algebra, and in the Friedmann limits $k = -1$ ($d = 0$) and $k = 0$ ($d = b + f = 0$) it becomes type V and I, respectively.

## VIII. Case 1.1.2.2 of Ref. 3, Bianchi type IX subcase.

The case 1.1.2.2 contains three different subcases that are of Bianchi types IX, VIII and VII$_0$. The type IX subcase requires some adaptation of the formulae given in Ref. 3.

For type IX, $g/c > 0$. Then, as seen from eq. (5.16) in Ref. 3, $B/c < 0$, or else (5.16) would lead to a contradiction. These two inequalities imply that $gB < 0$, while eqs. (5.26) and (5.27) in Ref. 3 are adapted to the case $gB > 0$. Hence, a re-adaptation of these formulae to type IX is necessary first. We define:
$$B := -\overline{B}, \qquad \lambda = i\overline{\lambda}, \qquad k_{12} = i\overline{k_{12}}, \qquad k_{23} = i\overline{k_{23}} \tag{8.1}$$

(the overbars simply denote new symbols that will be real), so that instead of (5.16), (5.23), (5.26) and (5.27) from Ref. 3, we obtain:
$$K = \frac{1}{2D}\left(\frac{2\overline{B} - gy^2}{c}\right)^{1/2}, \qquad \delta^2 := (\overline{B}/D)^2 + (2c\gamma)^2, \qquad \overline{\lambda}^2 := \frac{g\overline{B}}{8\delta^4 D^2},$$

$$R = 2cD^2 y/(\overline{B}K), \qquad \int K^{-3} R \mathrm{d}y = \frac{4c^2 D^4}{g\overline{B}K^2},$$

$$v = \overline{B}t + 2cD\gamma z, \qquad U = h_{12}\sin(2\overline{\lambda}v) + \overline{k_{12}}\cos(2\overline{\lambda}v); \tag{8.2}$$

$$g_{11} = y^2 + K^2 H_{11} + 4\frac{c\gamma D}{\overline{B}} y K H_{13} + 8\frac{(c\gamma D)^2}{g\overline{B}} H_{33},$$

$$g_{12} = H_{12} + 2\frac{c\gamma D}{\overline{B}K} y H_{23}, \qquad g_{13} = K H_{13} + 2\frac{c\gamma D}{\overline{B}} y H_{33},$$

$$g_{22} = H_{22}/K^2, \qquad g_{23} = H_{23}/K, \qquad g_{33} = H_{33}; \tag{8.3}$$

$$H_{11} = -\frac{cD^2}{2\delta^2\overline{\lambda}} U + h_{11}, \qquad H_{12} = h_{12}\cos(2\overline{\lambda}v) - \overline{k_{12}}\sin(2\overline{\lambda}v),$$

$$H_{13} = -\frac{cD^2}{2\delta^2\overline{\lambda}}[h_{23}\sin(\overline{\lambda}v) + \overline{k_{23}}\cos(\overline{\lambda}v)],$$

$$H_{22} = 2\frac{\delta^2\overline{\lambda}}{cD^2} U + \frac{g\overline{B}}{2c^2 D^6} h_{11} + \frac{8c\gamma^2}{\overline{B}D^2} h_{33},$$

$$H_{23} = h_{23}\cos(\overline{\lambda}v) - \overline{k_{23}}\sin(\overline{\lambda}v), \qquad H_{33} = h_{33}, \tag{8.4}$$

where $\overline{B}$, $c$, $D$, $g$ and $\gamma$ are arbitrary constants, and all the $h_{ij}$, $\overline{k_{12}}$ and $\overline{k_{23}}$ are arbitrary functions of the argument:

$$T = t - \frac{\overline{B}}{2cD\gamma} z. \tag{8.5}$$

Eqs. (8.2) – (8.5) are written in the Plebański coordinates.

The Killing fields for the metric (8.3) – (8.4) are:

$$k^{\alpha}_{(1)} = \delta^{\alpha}_1,$$

$$k^{\alpha}_{(2)} = \cos(Dx/2)\left[(K - yK_{,y})\delta^{\alpha}_0 + K_{,y}\delta^{\alpha}_1 + \frac{\gamma}{DK}\delta^{\alpha}_3\right] + \frac{1}{2} DK \sin(Dx/2)\delta^{\alpha}_2,$$

$$k^{\alpha}_{(3)} = \sin(Dx/2)\left[(K - yK_{,y})\delta^{\alpha}_0 + K_{,y}\delta^{\alpha}_1 + \frac{\gamma}{DK}\delta^{\alpha}_3\right] - \frac{1}{2} DK \cos(Dx/2)\delta^{\alpha}_2. \tag{8.6}$$

(Note: the first commutator in eq. (5.20) in Ref. 3 should have a minus on the right-hand side.) The analog of (5.1) here is:

$$\cos(Dx/2)k^{\alpha}_{(2)} + \sin(Dx/2)k^{\alpha}_{(3)} - K_{,y} k^{\alpha}_{(1)} = (K - yK_{,y})u^{\alpha} + +[\gamma/(DKn)]w^{\alpha}. \tag{8.7}$$

In agreement with this, the King–Ellis measure of tilt is here:

$$\sqrt{-g} N_{\alpha} u^{\alpha} = \gamma/2. \tag{8.8}$$

Note that:

$$K_{,y} = -\frac{gy}{4cD^2 K}, \qquad K - yK_{,y} = \frac{\overline{B}}{2cD^2 K}, \tag{8.9}$$

The case presently considered is the only one of type IX in the whole classification. Therefore:

1. This is the only place where the $k = +1$ Friedmann model will appear as a limit.
2. The models represented by eqs. (8.2) – (8.6) include those considered by Gödel[10]. (Ours are in fact more general because the tilt of the symmetry orbits with respect to the velocity field is an arbitrary parameter here.) We shall deal with this point further on.

For later considerations, it will be convenient to reparametrize the metric (8.4) once more, as follows:

$$\overline{G} := \left(h_{12}{}^2 + \overline{k_{12}}^2\right)^{1/2}, \qquad h_{12} = -\overline{G}\sin(2\beta), \qquad \overline{k_{12}} = \overline{G}\cos(2\beta),$$

$$F := \left(h_{23}{}^2 + \overline{k_{23}}^2\right)^{1/2}, \qquad h_{23} = F\cos(\alpha), \qquad \overline{k_{23}} = F\sin(\alpha), \qquad (8.10)$$

where $\overline{G}$, $F$, $\alpha$ and $\beta$ are new functions of the $T$ given by (8.3); and also to transform the coordinate $y$ by:

$$y = \sqrt{2\overline{B}/g}\cos(\vartheta). \qquad (8.11)$$

From now on, the $x^2$-coordinate will be $\vartheta$. Then, in order to set the rotation and the tilt to zero, the following further reparametrization is needed:

$$\overline{B} = b\omega_0{}^2, \qquad \gamma = h\omega_0, \qquad \overline{G} = G/\omega_0, \qquad h_{11} = G_{11}/\omega_0{}^2. \qquad (8.12)$$

Let us note that:

$$\overline{\lambda}v = \frac{\sqrt{bg}D(\omega_0 bt + 2cDhz)}{2\sqrt{2}[(\omega_0 b)^2 + (2cDh)^2]} \xrightarrow[\omega_0 \to 0]{} \frac{\sqrt{bg}}{4\sqrt{2}ch}z. \qquad (8.13)$$

The metric (8.5), reparametrized by (8.10) – (8.12), becomes:

$$g_{11} = \frac{2b}{g}\cos^2\vartheta\,\omega_0{}^2 + \frac{2b}{4cD^2}G_{11}\sin^2\vartheta - 4\frac{chD^3}{g}\sqrt{\frac{2c}{b}}F\sin(\overline{\lambda}v + \alpha)\sin\vartheta\cos\vartheta$$

$$-D\sqrt{\frac{b}{2g}}G\cos(2\overline{\lambda}v + 2\beta)\sin^2\vartheta + 8\frac{(chD)^2}{bg}h_{33},$$

$$g_{12} = \sqrt{2b/g}\,G\sin(2\overline{\lambda}v + 2\beta)\sin\vartheta - 4\sqrt{2}(chD^2/g)\sqrt{c/b}\,F\cos(\overline{\lambda}v + \alpha)\cos\vartheta$$

$$g_{13} = 2\sqrt{2}\frac{chD}{\sqrt{bg}}h_{33}\cos\vartheta - D^2\sqrt{\frac{c}{g}}F\sin(\overline{\lambda}v + \alpha)\sin\vartheta,$$

$$g_{22} = 2\frac{b}{cD^4}G_{11} + 32\frac{(ch)^2}{bg}h_{33} + 2\sqrt{2}\frac{1}{D}\sqrt{\frac{b}{g}}G\cos(2\overline{\lambda}v + 2\beta),$$

$$g_{23} = -2D\sqrt{\frac{c}{g}}F\cos(\overline{\lambda}v + \alpha), \qquad g_{33} = h_{33}. \qquad (8.14)$$

In the limit $\omega_0 \to 0$, the velocity field $u^\alpha = \delta^\alpha{}_0$ will have zero shear when:

$$\alpha = \text{const}, \qquad \beta = \text{const},$$

$$(G, F, G_{11}, h_{33}) = -(C_{12}, C_{23}, C_{11}, 1)R^2(t). \qquad (8.15)$$

With use of (8.13) it may be verified now that the $k = +1$ Friedmann limit will result from (8.14) – (8.15) when $\omega_0 \to 0$ and:

$$C_{12} = C_{23} = C_{11} = 0. \tag{8.16}$$

The resulting representation of the Friedmann model (again an exotic one) is identical, up to rescalings of coordinates, to the one derived by Behr[11]

$$\mathrm{d}s^2 = \mathrm{d}t^2 - R^2(t)\left[8\frac{(chD)^2}{bg}\mathrm{d}x^2 + 4\sqrt{2}\frac{chD}{\sqrt{bg}}\cos\vartheta\mathrm{d}x\mathrm{d}z + 32\frac{(ch)^2}{bg}\mathrm{d}\vartheta^2 + \mathrm{d}z^2\right]. \tag{8.17}$$

The $k = 0$ Friedmann model will result from this after the transformation/reparametrization:

$$\vartheta = \arccos(ky), \qquad h = H/k, \qquad x = kx', \tag{8.18}$$

in the limit $k \to 0$.

For the $k = +1$ Friedmann limit, the algebra of the Killing fields (8.6), suitably transformed by use of (8.9), (8.11) and (8.12), is still of Bianchi type IX. For the $k = 0$ limit, the algebra $\{l_{(1)}, l_{(2)}, l_{(3)}\} := k\{k_{(1)}, k_{(2)}, k_{(3)}\}$ is of Bianchi type I when $k \to 0$.

As stated above, the class of models defined by (8.2) – (8.5) must contain the one considered by Gödel in Ref. 10. This is so because two of Gödel's assumptions (dust source and nonzero rotation) place his class within our collection, and the third assumption (compact spaces $t = \mathrm{const}$, i.e. Bianchi type IX; the Bianchi classification and terminology had not yet been in common use in Gödel's time) uniquely points to the subcase I of our case 1.1.2.2. Gödel presented several properties of these models in the form of theorems, but mostly without proofs and almost without formulae. It would be an interesting exercise to see how Gödel's theorems apply to the explicitly given metric (8.2) – (8.5).

In particular, one of his statements seems to need a refinement. He said that there exist $\infty^8$ rotating solutions satisfying all his requirements. This means that the collection of all solutions of the Einstein equations for (8.2) – (8.5) should be labeled by 8 arbitrary constants. One can understand how this happens from Ref. 4, where the Einstein equations were investigated for an equally general Bianchi type V class. Of the 6 unknown functions in the initial metric, one ($h_{33}$ in Ref. 4) is determined by an algebraic relation, two of the Einstein equations are of first order and can be used to eliminate two more functions, and then the remaining 3 functions obey equations of second order. This gives 8 constants indeed. However, the tilt parameter ($\gamma$ in (8.7) – (8.8)) is one more arbitrary constant that is contained in the metric even before the Einstein equations are considered.

Rotating dust models of Bianchi type IX were considered by Behr [11], with simplifying assumptions about the metric. Similarly as in Ref. 4, the main conclusion seems to be that whatever one does with the Einstein equations, no solution comes within sight.

## IX. Case 1.1.2.2 of Ref. 3, Bianchi types VIII and VII$_0$.

The subcase of case 1.1.2.2 that corresponds to the Bianchi type VIII is defined by:

$$g/c < 0 \tag{9.1}$$

in eqs. (5.16) – (5.27) in Ref. 3. Then, $B/c$ and, consequently, $Bg$ can have any sign at this point. Only the $k = 0$ Friedmann model can be contained as a subcase here.

The cases $Bg \ne 0$ and $Bg = 0$ have to be considered separately. When $Bg \ne 0$, we take eqs. (5.23) – (5.28) in Ref. 3 with the following specializations:

$$h_{12} = k_{12} = h_{23} = k_{23} = 0,$$

$$B = b\omega_0^2, \qquad (\gamma, y) = (h, \tilde{y})\omega_0, \qquad h_{11} = G_{11}/\omega_0^2. \tag{9.2}$$

Then:

$$K = \omega_0 \tilde{K}, \qquad \tilde{K} = \frac{1}{2D}\sqrt{-\frac{g\tilde{y}^2 + 2b}{c}},$$

$$g_{11} = (\omega_0 \tilde{y})^2 - 8\frac{(cDh)^2}{bg}h_{33} + \tilde{K}^2 G_{11},$$

$$g_{12} = 0, \qquad g_{13} = -2(cDh/b)\tilde{y}h_{33},$$

$$g_{22} = \tilde{K}^{-2}\left(-\frac{bg}{2c^2 D^6}G_{11} - 8\frac{ch^2}{bD^2}h_{33}\right), \qquad g_{23} = 0, \qquad g_{33} = h_{33}. \tag{9.3}$$

It is now seen that the proper signature will result only when:

$$b/c > 0. \tag{9.4}$$

In order to obtain the $k = 0$ Friedmann model from (9.3), we then rescale the constants again as follows:

$$b = b_0\sqrt{g}, \qquad D = dg^{1/4}, \qquad h = Hg^{3/4}, \tag{9.5}$$

and take the limit $g \to 0$. The limiting metric is:

$$ds^2 = dt^2 + G_{11}\left(\frac{b_0}{2cd^2}dx^2 + \frac{1}{d^4}d\tilde{y}^2\right) + h_{33}dz^2 \tag{9.6}$$

and in the limit of zero shear, $G_{11} = C_{11}h_{33} = -C_{11}R^2(t)$, this becomes the $k = 0$ Friedmann models indeed.

The Killing fields for this case are given by (8.6). After the rescalings (9.2) and (9.5), the following basis of the symmetry algebra is obtained:

$$k^\alpha_{(1)} = \delta^\alpha{}_1,$$

$$l^\alpha_{(2)} = \sqrt{\frac{b_0}{2cg}}\frac{1}{H}k^\alpha_{(3)} \xrightarrow[g\to 0]{} \frac{1}{4cH}\left(-\frac{\tilde{y}}{d}\delta^\alpha{}_1 + \frac{1}{2}b_0 dx\delta^\alpha{}_2\right) + \delta^\alpha{}_3,$$

$$l^\alpha_{(3)} = -2\sqrt{\frac{2c}{b_0}}g^{-1/4}k^\alpha_{(3)} \xrightarrow[g\to 0]{} \delta^\alpha{}_2. \tag{9.7}$$

This is of Bianchi type $VII_0$.

When $Bg = 0$, and the Bianchi type is VIII, we must have:

$$B = 0 \ne g. \tag{9.8}$$

The metric is then found from eqs. (5.28) and (5.23) – (5.26) in Ref. 3, suitably adapted. With $B = 0$, the arbitrary functions depend only on $t$. The metric needs then to be rescaled as follows:

$$(\gamma, y) = (h, \tilde{y})\omega_0, \qquad (h_{12}, h_{13}, h_{23}) = (G_{12}, G_{13}, G_{23})/\omega_0, \qquad h_{11} = G_{11}/\omega_0{}^2 \qquad (9.9)$$

and the result of the rescaling is:

$$g_{11} = (\omega_0 \tilde{y})^2 - \frac{g\tilde{y}^2}{4cD^2}G_{11} + \frac{Dg}{8ch}\tilde{y}^2 z G_{12} - 2\sqrt{-c/g}hG_{13}$$

$$-\frac{g^2}{2^5 c^2 h}\sqrt{-c/g}(\tilde{y}z)^2 G_{13} - \frac{D^4 g}{2^6 ch^2}(\tilde{y}z)^2 h_{22} + \frac{1}{4}D^2 \tilde{y}z G_{23} + \left(\frac{Dg}{2^4 ch}\right)^2 (\tilde{y}z)^3 G_{23}$$

$$+4\left(\frac{cDh}{g\tilde{y}}\right)^2 h_{33} + \frac{1}{8}(Dz)^2 h_{33} + \left(\frac{Dg}{2^5 ch}\tilde{y}z^2\right)^2 h_{33},$$

$$g_{12} = G_{12} - \frac{g}{2cD}\sqrt{-c/g}zG_{13} - \frac{D^3}{4h}zh_{22} + 2\frac{cDh}{g\tilde{y}}G_{23} + 3\frac{Dg}{2^5 ch}\tilde{y}z^2 G_{23}$$

$$+2\frac{cDh}{g\tilde{y}^2}zh_{33} + \frac{Dg}{2^5 ch}z^3 h_{33},$$

$$g_{13} = \frac{1}{2D}\sqrt{-g/c}\tilde{y}G_{13} + \frac{Dg}{2^5 ch}\tilde{y}^2 z G_{23} + 2\frac{cDh}{g\tilde{y}}h_{33} + \frac{Dg}{2^5 ch}\tilde{y}z^2 h_{33},$$

$$g_{22} = -4\frac{cD^2}{g\tilde{y}^2}h_{22} + 2(z/\tilde{y})G_{23} + (z/\tilde{y})^2 h_{33}$$

$$g_{23} = G_{23} + (z/\tilde{y})h_{33}, \qquad g_{33} = h_{33}. \qquad (9.10)$$

In the limit $\omega_0 \to 0$ one term in $g_{11}$ disappears and the $h_{ij}$ depend only on $t$. The shearfree limit is then attained when:

$$G_{ij} = -C_{ij}R^2(t), \qquad h_{33} = -R^2(t), \qquad h_{22} = -C_{22}R^2(t). \qquad (9.11)$$

To find the Friedmann limit we then assume that:

$$C_{12} = C_{13} = C_{23} = 0, \qquad h = HD, \qquad g = -cG^2 D^2, \qquad (9.12)$$

where $H$ and $G$ are new constants (the last definition takes into acccount that $g/c < 0$ in type VIII), and let $\omega_0 \to 0$, $D \to 0$. The resulting metric is:

$$ds^2 = dt^2 - C_{11}\left(\frac{1}{2}G\tilde{y}R\right)^2 dx^2 - C_{22}\left(2\frac{R}{G\tilde{y}}\right)^2 d\tilde{y}^2 - R^2\left(-2\frac{H}{G^2\tilde{y}}dx + \frac{z}{\tilde{y}}d\tilde{y} + dz\right)^2. \qquad (9.13)$$

The $k = 0$ Friedmann limit results from this when:

$$\tilde{y} = e^{ku}, \qquad C_{22} = (D_{22}/k)^2, \qquad k \to 0, \qquad (9.14)$$

With (9.8), (9.9) and (9.12), the Killing fields become:

$$k_{(1)}^\alpha = \delta_1^\alpha,$$

$$k_{(2)}^\alpha = \frac{1}{2}G\cos(Dx/2)\delta_1^\alpha + \frac{1}{4}GD\tilde{y}\sin(Dx/2)\delta_2^\alpha + \frac{2H}{G\tilde{y}}\cos(Dx/2)\delta_3^\alpha,$$

$$k_{(3)}^\alpha = \frac{1}{2}G\sin(Dx/2)\delta_1^\alpha - \frac{1}{4}GD\tilde{y}\cos(Dx/2)\delta_2^\alpha + \frac{2H}{G\tilde{y}}\sin(Dx/2)\delta_3^\alpha. \qquad (9.15)$$

Before the limit $D \to 0$ can be taken, $k_{(3)}$ needs to be redefined by $k'_{(3)} = (1/D)k_{(3)}$. The basis in the limit becomes:

$$k_{(1)}^\alpha = \delta_1^\alpha, \qquad k_{(2)}^\alpha = \frac{1}{2}G\delta_1^\alpha + \frac{2H}{G\tilde{y}}\delta_3^\alpha,$$

$$k'^\alpha_{(3)} = \frac{1}{4}Gx\delta_1^\alpha - \frac{1}{4}G\tilde{y}\delta_2^\alpha + \frac{Hx}{G\tilde{y}}\delta_3^\alpha. \qquad (9.16)$$

This is of type VI$_0$. We now transform $\tilde{y}$ by (9.10), and redefine $k'_{(3)}$ once more:

$$l_{(3)}^\alpha = -(4k/G)k'^\alpha_{(3)}. \qquad (9.17)$$

In the limit $k \to 0$, the following basis then results:

$$k_{(1)}^\alpha = \delta_1^\alpha, \qquad k_{(2)}^\alpha = \frac{1}{2}G\delta_1^\alpha + \frac{2H}{G}\delta_3^\alpha, \qquad l_{(3)}^\alpha = \delta_2^\alpha \qquad (9.18)$$

($u$ being now the $x^2$), which is clearly of Bianchi type I.

Finally, when the Bianchi type is VII$_0$ (i.e. $g = 0$), the metric results by a simple specialization of eqs. (5.23) – (5.25) and (5.28) in Ref. 3. In this case necessarily $b/c < 0$ and:

$$K = \frac{1}{D}\sqrt{\frac{-b}{2c}} = \text{const.} \qquad (9.19)$$

The rescaling that will allow to calculate the limit $\omega_0 \to 0$ is:

$$(B, y) = (b, \tilde{y})\omega_0^2, \qquad D = d\omega_0,$$

$$(h_{12}, h_{23}) = (G_{12}, G_{23})/\omega_0, \qquad h_{22} = G_{22}/\omega_0^2. \qquad (9.20)$$

The argument of the arbitrary functions must then be redefined so that it becomes:

$$\tilde{u} = u/(2cD\gamma) = t + \frac{B}{2cD\gamma}z \xrightarrow[\omega_0 \to 0]{} t. \qquad (9.21)$$

The limit $\omega \to 0$ of the metric is then:

$$ds^2 = dt^2 - \frac{b}{2cd^2}h_{11}dx^2 - 2G_{12}dxd\tilde{y} + \frac{2}{d}\sqrt{\frac{-b}{2c}}h_{13}dxdz - 2\frac{cd^2}{b}C_{22}d\tilde{y}^2 + 2G_{23}d\tilde{y}dz + h_{33}dz^2. \qquad (9.22)$$

The $k = 0$ Friedmann limit results from here when shear is set to zero, i.e. when $g_{ij} = -C_{ij}R^2(t)$.

The basis of the Killing fields in the limit $\omega_0 \to 0$ is found as follows:

$$k_{(1)}^\alpha = \delta_1^\alpha, \qquad l_{(2)}^\alpha = \lim_{\omega_0 \to 0}\left(\frac{K\omega_0}{d\gamma}k_2^\alpha\right) = \delta_3^\alpha, \qquad l_{(3)}^\alpha = \lim_{\omega_0 \to 0}\left(-\frac{2}{dK}\omega_0 k_3^\alpha\right) = \delta_2^\alpha. \qquad (9.23)$$

# X. Cases 2.1 of Ref. 3.

In the case 2.1.1 the transformation back to the Plebański coordinates is the inverse of (7.16) in Ref. 3, and when applied to (7.18) there, it gives the following metric:

$$g_{11} = y^2 - (y+V)^2 + h_{11} + (b+f)[(y+V)h_{12} + (\gamma/c)Vh_{13}]$$

$$+ \frac{1}{4}(b+f)^2[(y+V)^2 h_{22} + 2(\gamma/c)V(y+V)h_{23} + (\gamma V/c)^2 h_{33}],$$

$$g_{12} = \frac{2}{b+f}(y+V) - h_{12} - \frac{1}{2}(b+f)[(y+V)h_{22} + (\gamma/c)Vh_{23}],$$

$$g_{13} = h_{13} + \frac{1}{2}(b+f)[(y+V)h_{23} + (\gamma/c)Vh_{33}],$$

$$g_{22} = -4/(b+f)^2 + h_{22}, \qquad g_{23} = -h_{23}, \qquad g_{33} = h_{33}, \qquad (10.1)$$

where

$$V := \frac{1}{2}(b+f)t + y, \qquad (10.2)$$

and the arbitrary functions $h_{ij}$ depend on

$$T := t + \frac{2y}{b+f} + \frac{2c}{\gamma(b+f)} z. \qquad (10.3)$$

(Note two typos in Ref. 3: in eq. (7.18), the coefficient of $Wh_{23}$ in $g_{22}$ is $2\gamma$, not $2b$, and in (7.17), in the formula for $u^\alpha$, there should be a $(W/c)$ in front of $\delta^\alpha{}_0$.) The Killing fields for this metric are:

$$k_{(1)}^\alpha = \delta_1^\alpha, \qquad k_{(2)}^\alpha = e^{(b+f)x/2}\left\{cx\delta_0^\alpha - c[1 + \frac{1}{2}(b+f)x]\delta_2^\alpha + \gamma\delta_3^\alpha\right\}$$

$$k_{(3)}^\alpha = e^{(b+f)x/2}[\delta_0^\alpha - \frac{1}{2}(b+f)\delta_2^\alpha], \qquad (10.4)$$

and they form a Bianchi type IV algebra.

The analog of (5.5) is

$$k_{(2)}^\alpha - \frac{2c}{b+f}[1 + \frac{1}{2}(b+f)x]k_{(3)}^\alpha = e^{(b+f)x/2}\left(\frac{2c}{b+f}u^\alpha + \frac{\gamma}{n}w^\alpha\right), \qquad (10.5)$$

and the King-Ellis measure of tilt is

$$\sqrt{-g}\,u^\alpha N_\alpha = -\frac{1}{2}(b+f)\gamma e^{(b+f)x}. \qquad (10.6)$$

The redefinitions needed to make the limit $\omega_0 \to 0$ finite are

$$(y, c) = \omega_0(\tilde{y}, C)$$

$$h_{11} = H_{11} - \frac{1}{2}(b+f)^2 T[h_{12} + (\gamma/c)h_{13}]$$

$$-\frac{1}{16}(b+f)^4 T^2 [H_{22}/\omega_0{}^2 + 2(\gamma/c)H_{23}/\omega_0 + (\gamma/c)^2 h_{33}],$$

$$h_{12} = H_{12}/\omega_0 - \frac{1}{4}(b+f)^2 T [H_{22}/\omega_0{}^2 + (\gamma/c)H_{23}/\omega_0],$$

$$h_{13} = H_{13} - \frac{1}{4}(b+f)^2 T [H_{23}/\omega_0 + (\gamma/c)h_{33}],$$

$$h_{22} = H_{22}/\omega_0{}^2, \qquad h_{23} = H_{23}/\omega_0, \tag{10.7}$$

and the resulting metric is

$$g_{11} = (\omega_0 \tilde{y})^2 - \left[\frac{1}{2}(b+f)t + 2\omega_0 \tilde{y}\right]^2 + H_{11}$$

$$+ (b+f)[Y H_{12} - z H_{13}] + \frac{1}{4}(b+f)^2 [Y^2 H_{22} - 2YzH_{23} + z^2 H_{33}],$$

$$g_{12} = \omega_0 t + \frac{4\omega_0{}^2}{b+f} \tilde{y} - H_{12} - \frac{1}{2}(b+f)[Y H_{22} - z H_{23}],$$

$$g_{13} = H_{13} + \frac{1}{2}(b+f)[Y H_{23} - z H_{33}],$$

$$g_{22} = H_{22} - 4\omega_0{}^2/(b+f)^2, \qquad g_{23} = -H_{23}, \qquad g_{33} = h_{33} = H_{33},$$

$$Y := \tilde{y} - Cz/\gamma. \tag{10.8}$$

In the limit $\omega_0 \to 0$, all $H_{ij}$ become functions of $t$, and the Killing fields become

$$k_{(1)}^\alpha = \delta_1^\alpha,$$

$$l_{(2)}^\alpha = \lim_{\omega_0 \to 0} k_{(2)}^\alpha = e^{(b+f)x/2} \left\{ -C[1 + \frac{1}{2}(b+f)x]\delta_2^\alpha + \gamma \delta_3^\alpha \right\}$$

$$l_{(3)}^\alpha = \lim_{\omega_0 \to 0} \left( -\frac{2}{b+f} \omega_0 k_{(3)}^\alpha \right) = e^{(b+f)x/2} \delta_2^\alpha, \tag{10.9}$$

still of type IV.

The shearfree limit of (10.8) is

$$H_{11} = -C_{11} R^2(t) + \left[\frac{1}{2}(b+f)t\right]^2,$$

$$\text{other } H_{ij} = -C_{ij} R^2(t), \qquad C_{33} = 1. \tag{10.10}$$

The $k = -1$ Friedmann model will then result when $C = 0$ (and, consequently, $Y = \tilde{y}$), the $k = 0$ Friedmann model will result when $b + f = 0$, with no condition on $C$. Both limits can be easily taken also in the Killing fields (10.9), with $C = 0$ they become of type V, with $b + f = 0$ they become of type I.

The case 2.1.2 was shown in Ref. 3 to be included in 2.1.1 as a subcase.

# XI. Case 2.2.1.1 of Ref. 3.

This case includes two subcases, $\mathcal{A} \neq 0$ and $\mathcal{A} = 0$, given by eqs. (9.11) – (9.15) in Ref. 3. Both are of Bianchi type VIII. The coordinates used there are those of Plebański.

With $\mathcal{A} \neq 0$, eqs. (9.14) in Ref. 3 are adapted to the case $\mathcal{A} < 0$. However, when $\mathcal{A} < 0$, the limit of constant curvature in the spaces $t = $ const has a wrong signature. Therefore, the formulae must be re-adapted to $\mathcal{A} > 0$. This is the result:

$$g_{11} = y^2 \left( \frac{U}{\sqrt{2\mathcal{A}}} + \frac{\gamma V}{\mathcal{A}\sqrt{2\mathcal{A}}} + h_{11} \right),$$

$$g_{12} = h_{12} \cos(2\lambda v) - k_{12} \sin(2\lambda v) + \frac{\gamma}{2\mathcal{A}} \left[ h_{23} \cos(\lambda v) - k_{23} \sin(\lambda v) \right],$$

$$g_{13} = y \left( \frac{V}{\sqrt{2\mathcal{A}}} + \frac{\gamma}{2\mathcal{A}} h_{33} \right), \qquad g_{22} = y^{-2} \left( -\sqrt{2\mathcal{A}} U + 2\mathcal{A} h_{11} - \frac{\gamma^2}{2\mathcal{A}} h_{33} \right),$$

$$g_{23} = y^{-1} \left[ h_{23} \cos(\lambda v) - k_{23} \sin(\lambda v) \right], \qquad g_{33} = H_{33} = h_{33},$$

$$U := h_{12} \sin(2\lambda v) + k_{12} \cos(2\lambda v), \qquad V := h_{23} \sin(\lambda v) + k_{23} \cos(\lambda v),$$

$$\lambda^2 = 2\mathcal{A}/(4\mathcal{A}^2 + \gamma^2)^2, \qquad v = 2\mathcal{A}t + \gamma z, \tag{11.1}$$

and the $h_{ij}$ are arbitrary functions of

$$T = t - 2\mathcal{A}z/\gamma. \tag{11.2}$$

The Killing fields for (11.1) are

$$k_{(1)}^\alpha = \delta_1^\alpha, \qquad k_{(2)}^\alpha = (2\mathcal{A}/y)\delta_0^\alpha + (-\mathcal{A}/y^2 + x^2/2)\delta_1^\alpha - xy\delta_2^\alpha + (\gamma/y)\delta_3^\alpha,$$

$$k_{(3)}^\alpha = x\delta_1^\alpha - y\delta_2^\alpha, \tag{11.3}$$

The analog of (5.1) is:

$$(\mathcal{A}/y^2 + x^2/2)k_{(1)}^\alpha + k_{(2)}^\alpha - xk_{(3)}^\alpha = (2\mathcal{A}/y)u^\alpha + \frac{\gamma}{yn}w^\alpha, \tag{11.4}$$

and the King-Ellis measure of tilt is:

$$\sqrt{-g}u^\alpha N_\alpha = \gamma. \tag{11.5}$$

The redefinitions needed to make the limit $\omega_0 \to 0$ of (11.1) finite are:

$$(y, \gamma) = \omega_0(\tilde{y}, h), \qquad \mathcal{A} = \frac{1}{2}(a\omega_0)^2,$$

$$h_{11} = H_{11}/\omega_0^2, \qquad (h_{12}, k_{12}) = (H_{12}, K_{12})/\omega_0. \tag{11.6}$$

Note that with (11.6) we have

$$\lambda v \xrightarrow[\omega_0 \to 0]{} az/h. \tag{11.7}$$

The reparametrized metric is

$$g_{11} = \tilde{y}^2 \left\{ 2hV/a^3 + H_{11} + a^{-1}\left[H_{12}\sin(2\lambda v) + K_{12}\cos(2\lambda v)\right] \right\},$$

$$g_{12} = (h/a^2)\left[h_{23}\cos(\lambda v) - k_{23}\sin(\lambda v)\right] + H_{12}\cos(2\lambda v) - K_{12}\sin(2\lambda v),$$

$$g_{13} = (\tilde{y}/a)(V + hh_{33}/a),$$

$$g_{22} = \tilde{y}^{-2}\left\{-a\left[H_{12}\sin(2\lambda v) + K_{12}\cos(2\lambda v)\right] + a^2 H_{11} - (h/a)^2 H_{33}\right\},$$

$$g_{23} = \tilde{y}^{-1}\left[h_{23}\cos(\lambda v) - k_{23}\sin(\lambda v)\right], \qquad g_{33} = H_{33} = h_{33}. \tag{11.8}$$

In the limit $\omega_0 \to 0$, all the arbitrary functions will depend only on $t$.

The $k=0$ Friedmann limit follows from (11.8) when the following further specialization and transformation is made:

$$h_{33} = -R^2(t), \qquad H_{11} = -(C_{11} + h^2)R^2(t)/a^4,$$

$$H_{12} = K_{12} = h_{23} = k_{23} = 0, \qquad x = a^2 x', \qquad \tilde{y} = e^{au}, \tag{11.9}$$

and then the limit $a \to 0$ is taken. The metric becomes then

$$ds^2 = dt^2 - R^2\left[C_{11}\left(dx'^2 + du^2\right) + (gdx + dz)^2\right], \tag{11.10}$$

which is clearly the $k=0$ Friedmann model.

The rescaling (11.6), followed by $\omega_0 \to 0$, and the rescaling (11.9), followed by $a \to 0$ transform the Killing fields (11.3) into an almost-standard Bianchi type I basis ($k_{(1)}^\alpha$ has to be replaced by $l_{(1)}^\alpha = a^2 k_{(1)}^\alpha$, and $k_{(3)}^\alpha$ has to be replaced by $l_{(3)}^\alpha = a k_{(3)}^\alpha$ before taking the limit $a \to 0$).

The case $\mathcal{A} = 0$ is given by eqs. (9.11) and (9.15) in Ref. 3. The rescalings needed there are

$$y = \omega_0 \tilde{y}, \qquad h_{11} = G_{11}/\omega_0^2, \qquad (h_{12}, h_{13}) = (G_{12}, G_{13})/\omega_0. \tag{11.11}$$

The arbitrary functions $h_{ij}$ depend only on $t$ from the beginning. The limit $\omega_0 \to 0$ of the rescaled metric is

$$ds^2 = dt^2 + \tilde{y}^2 G_{11} dx^2 + 2(zG_{13} + G_{12})dxd\tilde{y} + 2\tilde{y}G_{13}dxdz$$

$$+\tilde{y}^{-2}(h_{22} + 2zh_{23} + z^2 h_{33})d\tilde{y}^2 + 2\tilde{y}^{-1}(h_{23} + zh_{33})d\tilde{y}dz + h_{33}dz^2. \tag{11.12}$$

The $k=0$ Friedmann model results now when

$$G_{12} = G_{13} = 0, \qquad G_{11} = -C_{11}R^2(t),$$

$$h_{ij} = -C_{ij}R^2, \qquad C_{33} = 1, \qquad C_{22} = 1/a^2,$$

$$\tilde{y} = e^{au}, \qquad a \to 0. \tag{11.13}$$

The Killing fields need not be reconsidered because $\mathcal{A} = 0$ is an allowed subcase for (11.3).

# XII. Cases 2.2.1.2 of Ref. 3.

In considering these cases, we first have to correct two errors. The first error is that the arbitrary constant $y_0$ actually must be equal to zero in all the formulae. The second error is that one subcase was overlooked – it needs special treatment and is not included in the formulae given in sec. X of Ref. 3. This special case is defined by

$$g = 0 \tag{12.1}$$

and consequently $\mu_1 = 0$ and $\mu_2 = j$. It is because of $\mu_1 = 0$ that some of the formulae do not apply to this case.

The conclusion that

$$a = 0, \quad c = 1 \tag{12.2}$$

can be achieved by a change of the basis of the Killing fields is still valid. With (12.1) and (12.2), the solutions of eqs. (10.2) and (10.3) in Ref. 3 are

$$P = -jy + M, \quad L_3 = \gamma, \tag{12.3}$$

where $j$, $M$ and $\gamma$ are arbitrary constants. The resulting Killing fields are (by (10.6) from Ref. 3):

$$k_{(1)}^\alpha = \delta_1^\alpha, \quad k_{(2)}^\alpha = Mx\delta_0^\alpha - jx\delta_1^\alpha + (jy - M)\delta_2^\alpha + \gamma x\delta_3^\alpha,$$
$$k_{(3)}^\alpha = M\delta_0^\alpha - j\delta_1^\alpha + \gamma\delta_3^\alpha, \tag{12.4}$$

and they form a Bianchi type III algebra.

The coordinates are still those of Plebański at this point, so $u^\alpha$ and $w^\alpha$ and $g_{0\alpha}$ have their standard forms. The solution of the Killing equations is

$$g_{11} = (2M/j)y - (M/j)^2 + (jy - M)^2 h_{11} + 2(\gamma/j)(jy - M)h_{13} + (\gamma/j)^2 h_{33},$$

$$g_{12} = h_{12} + \frac{\gamma h_{23}}{j(jy - M)}, \quad g_{13} = (jy - M)h_{13} + (\gamma/j)h_{33},$$

$$g_{22} = h_{22}/(jy - M)^2, \quad g_{23} = h_{23}/(jy - M), \quad g_{33} = h_{33}, \tag{12.5}$$

where the $h_{ij}$ are arbitrary functions of the argument

$$T = t - (M/\gamma)z. \tag{12.6}$$

The Killing fieds (12.4) are a subcase of the general expression that will apply to the whole case 2.2.1.2 collection. The analog of (5.1) will be given further on for the whole class.

The rescalings needed to find the nonrotating limit of (12.5) – (12.6) are

$$(y, M) = \omega_0(\tilde{y}, m), \quad h_{11} = H_{11}/\omega_0^2,$$

$$(h_{12}, h_{13}) = (H_{12}, H_{13})/\omega_0. \tag{12.7}$$

The rescaled metric is

$$g_{11} = (2m/j)\omega_0^2\tilde{y} - (m\omega_0/j)^2 + (j\tilde{y} - m)^2 H_{11} + 2(\gamma/j)(j\tilde{y} - m)H_{13} + (\gamma/j)^2 h_{33},$$

$$g_{12} = H_{12} + \frac{\gamma h_{23}}{j(j\tilde{y} - m)}, \qquad g_{13} = (j\tilde{y} - m)H_{13} + (\gamma/j)h_{33},$$

$$g_{22} = H_{22}/(j\tilde{y} - m)^2, \qquad g_{23} = h_{23}/(j\tilde{y} - m), \qquad g_{33} = h_{33}, \qquad (12.8)$$

The $k = 0$ Friedmann limit is now obtained from (12.8) when shear is set to zero ($h_{ij} = -C_{ij}R^2(t)$, $C_{33} = 1$), and in addition

$$\gamma = hj, \qquad j \to 0. \qquad (12.9)$$

In order to make the limits $\omega_0 \to 0$ and $j \to 0$ compatible, it has to be assumed that $j \propto \omega_0{}^\alpha$, where $0 < \alpha < 1$, e.g. $\alpha = 1/2$. The Killing fields (12.4) become then an almost-standard Bianchi type I basis in the limit $\omega_0 \to 0$, but $k_{(3)}^\alpha$ has to be replaced by $l_{(3)}^\alpha = \omega_0{}^{-1/2}k_{(3)}^\alpha$.

The case 2.2.1.2 consists of 3 subcases, each of a different Bianchi type. The subcase $g < j^2/4$ is of Bianchi type VI$_h$, with the free parameter $j/(j^2 - 4g)^{1/2}$. However, the parametrization of the metric used in Ref. 3 is inconvenient for calculating the Friedmann limit. It will be more convenient to rewrite it in the parametrization in which $\mu_1$ and $\mu_2$ appear symmetrically. Therefore, instead of (10.15) – (10.17) from Ref. 3, we will use the following formulae:

$$U := M\cosh(DY) + N\sinh(DY), \qquad V := M\sinh(DY) + N\cosh(DY),$$

$$D := (j^2/4 - g)^{1/2} \qquad \Gamma := \frac{\gamma}{D(M^2 - N^2)},$$

$$P = e^{-jY/2}U, \qquad y = -\frac{j}{2g}P - \frac{D}{g}e^{-jY/2}V,$$

$$g_{11} = y^2 + h_{11}P^2 + 2\Gamma e^{-jY}UVh_{13} + (\gamma\Gamma/D)e^{-jY}h_{33},$$

$$g_{12} = h_{12} + \Gamma(V/U)h_{23}, \qquad g_{13} = h_{13}P + \Gamma e^{-jY/2}Vh_{33},$$

$$g_{22} = h_{22}/P^2, \qquad g_{23} = h_{23}/P, \qquad g_{33} = h_{33}, \qquad (12.10)$$

where $j$, $g$, $M$, $N$ and $\gamma$ are arbitrary constants, $y$ is one of the coordinates, $Y$ is just a parameter used to represent $P(y)$, and $h_{ij}$ are arbitrary functions of the coordinate $z$. Since the coordinates used in (12.10) are not those of Plebański, the $u^\alpha$ and $g_{0\alpha}$ do not have their standard forms, they are given by eqs. (10.9) in Ref. 3. The transformation back to the Plebański coordinates is the inverse of (10.8) in Ref. 3, after which we obtain

$$g_{12} = gth_{11} + h_{12} + \Gamma gt(V/U - \frac{1}{2}j/D)h_{13} + \Gamma(V/U)h_{23} - \Gamma^2 gt\left(1 + \frac{jV}{2DU}\right)h_{33},$$

$$g_{22} = P^{-2}\left\{(gt)^2 h_{11} + 2gth_{12} - (j\Gamma g/D)t(gth_{13} + h_{23}) + h_{22} + g(\Gamma gt/D)^2 h_{33}\right\},$$

$$g_{23} = (1/P)\left\{gth_{13} + h_{23} - \frac{j\Gamma gt}{2D}h_{33}\right\}, \qquad (12.11)$$

with $g_{11}$, $g_{13}$ and $g_{33}$ being the same as in (12.10). The $h_{ij}$ depend now on

$$T = t + \frac{D}{\Gamma g}z. \qquad (12.12)$$

The Killing fields corresponding to (12.11) – (12.12) (and to all the other subcases of case (2.2.1.2)) are

$$k^{\alpha}_{(1)} = \delta^{\alpha}_1, \qquad k^{\alpha}_{(2)} = x(P - yP_{,y})\delta^{\alpha}_0 + xP_{,y}\delta^{\alpha}_1 - P\delta^{\alpha}_2 + x(\gamma/P)e^{-jY}\delta^{\alpha}_3,$$

$$k^{\alpha}_{(3)} = (P - yP_{,y})\delta^{\alpha}_0 + P_{,y}\delta^{\alpha}_1 + (\gamma/P)e^{-jY}\delta^{\alpha}_3. \qquad (12.13)$$

The analog of (5.1) (again valid for all the subcases) is

$$k^{\alpha}_{(3)} - P_{,y}k^{\alpha}_{(1)} = (P - yP_{,y})u^{\alpha} + \frac{\gamma}{nP}e^{-jY}w^{\alpha}, \qquad (12.14)$$

and the King-Ellis measure of tilt is:

$$\sqrt{-g}u^{\alpha}N_{\alpha} = -\gamma e^{-jY}. \qquad (12.15)$$

The rescalings needed to make the limit $\omega_0 \to 0$ finite are

$$(M, N, \gamma) = (m, n, h)\omega_0, \qquad h_{11} = H_{11}/\omega_0^2, \qquad h_{13} = H_{13}/\omega_0. \qquad (12.16)$$

In consequence of this we have

$$(y, P, U, V) = (\tilde{y}, \tilde{P}, \tilde{U}, \tilde{V})\omega_0, \qquad (12.17)$$

where the symbols with a tilde are obtained from those on the left by replacing $(m, N) \to (m, n)$, and they do not depend on $\omega_0$. Also, from now on $Y$ will be used as the $x^2$-coordinate in place of $y$, so

$$dy = \omega_0 \tilde{P} dY. \qquad (12.18)$$

The rescalings (12.16) have to be accompanied by the following redefinitions of other functions in the metric:

$$h_{12} = -gTH_{11}/\omega_0^2 + H_{12}/\omega_0 + \frac{j\Gamma gT}{2D\omega_0}H_{13} + \Gamma^2 gTh_{33},$$

$$h_{22} = -(gT)^2 H_{11}/\omega_0^2 - 2gTh_{12} + H_{22} + \frac{j\Gamma g}{D}\left(gT^2 H_{13}/\omega_0 + Th_{23}\right) - g(\Gamma gT/D)^2 h_{33},$$

$$h_{23} = -gTH_{13}/\omega_0 + H_{23} + \frac{j\Gamma gT}{2D}h_{33}. \qquad (12.19)$$

The metric resulting after the redefinitions and the coordinate transformation is

$$\tilde{\Gamma} := \frac{h}{D(m^2 - n^2)},$$

$$g_{11} = (\omega_0 \tilde{y})^2 + H_{11}\tilde{P}^2 + \tilde{\Gamma}e^{-jY}\left[2\tilde{U}\tilde{V}H_{13} - (h/D)h_{33}\right],$$

$$g_{12} = \tilde{P}\left\{H_{12} - (D/\tilde{\Gamma})zH_{11} + (j/2 - D\tilde{V}/\tilde{U})zH_{13}\right.$$

$$\left. +\tilde{\Gamma}\left[(\tilde{V}/\tilde{U})H_{23} + (D + \frac{1}{2}j\tilde{V}/\tilde{U})zh_{33}\right]\right\},$$

$$g_{13} = H_{13}\tilde{P} + \tilde{\Gamma}e^{-jY/2}\tilde{V}h_{33},$$

$$g_{22} = (Dz/\tilde{\Gamma})^2 H_{11} - (D/\tilde{\Gamma})(2zH_{12} + jz^2 H_{13}) + H_{22} + jzH_{23} + gz^2 h_{33},$$
$$g_{23} = -(D/\tilde{\Gamma})zH_{13} + H_{23} + (j/2)zh_{33}, \qquad g_{33} = h_{33} = H_{33}, \tag{12.20}$$

In the limit $\omega_0 \to 0$, all the $H_{ij}$ will depend only on $t$.

The limit of zero shear is then, as usual, $H_{ij} = -C_{ij} R^2(t)$, $C_{33} = 1$, and the $k = -1$ Friedmann limit results from (12.20) when, in addition

$$h = HD, \qquad C_{13} = 0, \qquad D \to 0. \tag{12.21}$$

With $D = 0$ we have $\tilde{U} = m$, $\tilde{V} = n$. The, again rather exotic, representation of the limiting Friedmann model is

$$\mathrm{d}s^2 = \mathrm{d}t^2 - R^2(t) \left( mD_{11} e^{-jY/2} \mathrm{d}x + D_{12} \mathrm{d}Y \right)^2 - (D_{22} R \mathrm{d}Y)^2$$
$$- R^2 \left[ e^{-jY/2} \frac{Hn}{m^2 - n^2} \mathrm{d}x + (C_{23} + jz/2) \mathrm{d}Y + \mathrm{d}z \right]^2, \tag{12.22}$$

where

$$D_{11}^2 := C_{11}^2 - \left( \frac{H}{m^2 - n^2} \right)^2, \qquad D_{12} := C_{12}/D_{11},$$
$$D_{22}^2 := C_{22} - C_{23}^2 - D_{12}^2. \tag{12.23}$$

The $k = 0$ Friedmann limit results from (12.22) when $j = 0$.

The rescaling (12.16) and the limit $\omega_0 \to 0$ transform the Killing fields as follows

$$k_{(1)}^\alpha = \delta_1^\alpha, \qquad k_{(2)}^\alpha = (-j/2 + D\tilde{V}/\tilde{U})x\delta_1^\alpha - \delta_2^\alpha + (h/\tilde{U})x e^{-jY/2} \delta_3^\alpha,$$
$$k_{(3)}^\alpha = (-j/2 + D\tilde{V}/\tilde{U})\delta_1^\alpha + (h/\tilde{U}) e^{-jY/2} \delta_3^\alpha, \tag{12.24}$$

the algebra still being of type $VI_h$.

The further rescaling (12.21) and the limit $D \to 0$ transform (12.24) into a Bianchi type V algebra, but $k_{(3)}^\alpha$ has to be replaced by

$$l_{(3)}^\alpha = D^{-1} \left( k_{(3)}^\alpha + \frac{j}{2} k_{(1)}^\alpha \right) \xrightarrow[D \to 0]{} (n/m)\delta_1^\alpha + (H/m) e^{-jY/2} \delta_3^\alpha. \tag{12.25}$$

When $j = 0$ on top of $D \to 0$, the Bianchi type reduces to I.

The subcase with $g > j^2/4$ (Bianchi type $VII_h$ with the free parameter $j/(4g - j^2)^{1/2}$) is given by eqs. (10.18) – (10.19) in Ref. 3, with $y_0 = 0$. It is transformed back to the Plebański coordinates by the inverse of (10.8) there, and the result is very similar to our (12.10) – (12.11). Only the definitions of $U$, $V$ and $y$, and a few signs in the metric are different:

$$D = (g - j^2/4)^{1/2}, \qquad \Gamma := \frac{\gamma}{D(M^2 + N^2)},$$
$$U := M\cos(DY) + N\sin(DY), \qquad V := M\sin(DY) - N\cos(DY),$$
$$P = e^{-jY/2} U, \qquad y = -\frac{j}{2g} P + \frac{D}{g} e^{-jY/2} V,$$
$$g_{11} = y^2 + h_{11} P^2 + 2\Gamma e^{-jY} UV h_{13} + (\gamma\Gamma/D) e^{-jY} h_{33},$$

$$g_{12} = gth_{11} + h_{12} + \Gamma gt(V/U + \tfrac{1}{2}j/D)h_{13} + (\Gamma V/U)h_{23}$$
$$+\Gamma^2 gt\left(1 + \frac{jV}{2DU}\right)h_{33},$$
$$g_{13} = h_{13}P + \Gamma e^{-jY/2}Vh_{33},$$
$$g_{22} = P^{-2}\left[(gt)^2 h_{11} + 2gth_{12} + (j\Gamma/D)gt(gth_{13} + h_{23}) + h_{22} + g(\Gamma gt/D)^2 h_{33}\right],$$
$$g_{23} = P^{-1}\left(gth_{13} + h_{23} + \frac{j\Gamma gt}{2D}h_{33}\right), \qquad g_{33} = h_{33}. \tag{12.26}$$

The $h_{ij}$ are here functions of the argument
$$T = t - \frac{D}{\Gamma g}z. \tag{12.27}$$

The redefinitions in the constants and functions needed here are again (12.16) – (12.17) together with
$$h_{12} = -gTH_{11}/\omega_0^2 + H_{12}/\omega_0 - \frac{j\Gamma gT}{2D\omega_0}H_{13} - \Gamma^2 gTh_{33},$$
$$h_{22} = -(gT)^2 H_{11}/\omega_0^2 - 2gTh_{12} + H_{22} - (j\Gamma g/D)\left(gT^2 H_{13}/\omega_0 + Th_{23}\right) - g(\Gamma gT/D)^2 h_{33},$$
$$h_{23} = -gTH_{13}/\omega_0 + H_{23} - \frac{j\Gamma gT}{2D}h_{33}. \tag{12.28}$$

The metric resulting after all the redefinitions is
$$\tilde{\Gamma} := \frac{h}{D(m^2 + n^2)},$$
$$g_{11} = (\omega_0\tilde{y})^2 + H_{11}\tilde{P}^2 + \tilde{\Gamma}e^{-jY}\left[2\tilde{U}\tilde{V}H_{13} + (h/D)h_{33}\right],$$
$$g_{12} = \tilde{P}\Big[(D/\tilde{\Gamma})zH_{11} + H_{12} + (j/2 + D\tilde{V}/\tilde{U})zH_{13}$$
$$+ (\tilde{\Gamma}\tilde{V}/\tilde{U})H_{23} + \tilde{\Gamma}\left(D + \tfrac{1}{2}j\tilde{V}/\tilde{U}\right)zh_{33}\Big],$$
$$g_{13} = H_{13}\tilde{P} + \tilde{\Gamma}e^{-jY/2}\tilde{V}h_{33},$$
$$g_{22} = (Dz/\tilde{\Gamma})^2 H_{11} + (D/\tilde{\Gamma})(2zH_{12} + jz^2 H_{13}) + H_{22} + jzH_{23} + gz^2 h_{33},$$
$$g_{23} = (D/\tilde{\Gamma})zH_{13} + H_{23} + (j/2)zh_{33}, \qquad g_{33} = h_{33} = H_{33}, \tag{12.29}$$

Just as before, in the limit $\omega_0 \to 0$ the $H_{ij}$ will depend only on $t$, and the shearfree limit is found in the same way: $H_{ij} = -C_{ij}R^2$, $C_{33} = 1$.

The $k = -1$ Friedmann limit is now obtained in two ways: either
$$h = HD, \qquad D \to 0, \tag{12.30}$$
or
$$C_{11} = C_{13} = 0. \tag{12.31}$$

In the first case the Friedmann limit is

$$ds^2 = dt^2 - R^2(t)\left(mD_{11}e^{-jY/2}dx + D_{12}dY\right)^2 - (D_{22}RdY)^2$$

$$-R^2\left[e^{-jY/2}(mC_{13} + n\tilde{\Gamma})dx + (C_{23} + jz/2)dY + dz\right]^2, \tag{12.32}$$

where

$$D_{11}^2 := C_{11}^2 - C_{13}^2 + H^2/(m^2 + n^2)^2, \tag{12.33}$$

$D_{12}$ and $D_{22}$ being the same as in (12.23).

In the second case, the $k = -1$ Friedmann limit is

$$ds^2 = dt^2 - R^2(t)\left[D_{11}e^{-jY/2}\tilde{U}dx + (D_{12} + Dz)dY\right]^2 - (D_{22}RdY)^2$$

$$-R^2\left[e^{-jY/2}\tilde{V}\tilde{\Gamma}dx + (C_{23} + jz/2)dY + dz\right]^2, \tag{12.34}$$

where $D_{11}$ is defined as in (12.33), but with $C_{11} = C_{13} = 0$.

The Killing fields before redefinitions are still given by (12.13), but, in consequence of the different definitions of $P$ and $y$ in the present subcase, the Bianchi type is VII$_h$. In the Friedmann limit defined by (12.31), the Killing fields are transformed only by (12.16) – (12.17) followed by $\omega_0 \to 0$, and they still form a type VII$_h$ algebra. When (12.30) is imposed on top of (12.16) – (12.17) and $\omega_0 \to 0$, the Killing fields become the same as the limit $D \to 0$ of (12.24) – (12.25), i.e. the Bianchi type becomes V. This is an illustration of the fact, mentioned in sec. I, that the $k = -1$ Robertson-Walker geometry is a subcase of two Bianchi types simultaneously, they are exactly V and VII$_h$.

In both cases, the $k = 0$ Friedmann limit follows from (12.32) and (12.34) when $j = 0$. In the first case, the algebra of the Killing fields becomes type I, in the second case it becomes type VII$_0$, which is another illustration of the same kind of duality.

Finally, the third subcase of case 2.2.1.2 is given by eqs. (10.20) - (10.21) in Ref. 3, with $y_0 = 0$. There is one more typo there, the correct formula for $y$ is

$$y = -2P/j - 4Me^{-jY/2}/j^2. \tag{12.35}$$

This one is of Bianchi type IV. When transformed back to the Plebański coordinates (by the inverse of (10.8) in Ref. 3), it becomes

$$P = e^{-jY/2}(MY + N),$$

$$g_{11} = y^2 + h_{11}P^2 - 2(\gamma/M)e^{-jY/2}Ph_{13} + (\gamma/M)^2e^{-jY}h_{33},$$

$$g_{12} = (jt/2)^2h_{11} + h_{12} + \frac{j^3\gamma t}{8M^2}h_{13} - \frac{\gamma}{M(MY+N)}\left(\frac{1}{4}j^2th_{13} + h_{23} + \frac{j^3\gamma t}{8M^2}h_{33}\right),$$

$$g_{13} = h_{13}P - (\gamma/M)e^{-jY/2}h_{33},$$

$$g_{22} = P^{-2}\left[(\frac{1}{4}j^2t)^2h_{11} + \frac{1}{2}j^2th_{12} + \frac{j^5\gamma t^2}{16M^2}h_{13} + h_{22}\right.$$

$$\left. + \frac{j^3\gamma t}{4M^2}h_{23} + \left(\frac{j^3\gamma t}{8M^2}\right)^2 h_{33}\right],$$

$$g_{23} = P^{-1}\left(\frac{1}{4}j^2 t h_{13} + h_{23} + \frac{j^3\gamma t}{8M^2}h_{33}\right), \qquad g_{33} = h_{33}. \tag{12.36}$$

where the $h_{ij}$ are arbitrary functions of

$$T = t - \frac{4M^2}{j^2\gamma}z. \tag{12.37}$$

The redefinitions needed to calculate the limit $\omega_0 \to 0$ are

$$(M, N, \gamma) = (m, n, h)\omega_0, \qquad h_{11} = H_{11}/\omega_0^2, \qquad h_{13} = H_{13}/\omega_0,$$

$$h_{12} = -\frac{1}{4}j^2 T H_{11}/\omega_0^2 + H_{12}/\omega_0 - \frac{j^3 hT}{8(m\omega_0)^2}H_{13},$$

$$h_{22} = -(j^2 T/4)^2 H_{11}/\omega_0^2 - \frac{1}{2}j^2 T h_{12} - \frac{j^5 h T^2}{(4m\omega_0)^2}H_{13} + H_{22}$$

$$-\frac{j^3 hT}{4m^2\omega_0}h_{23} - \left(\frac{j^3 hT}{8m^2\omega_0}\right)^2 h_{33},$$

$$h_{23} = -\frac{1}{4}j^2 T H_{13}/\omega_0 + H_{23} - \frac{j^3 hT}{8m^2\omega_0}h_{33}. \tag{12.38}$$

We will denote, as before, $(y, P) = \omega_0(\tilde{y}, \tilde{P})$, and choose $Y$ as the new $x^2$-coordinate, so that $dy = \omega_0 \tilde{P} dY$. The metric that results is

$$g_{11} = (\omega_0 \tilde{y})^2 + H_{11}\tilde{P}^2 - 2(h/m)e^{-jY/2}\tilde{P}H_{13} + (h/m)^2 e^{-jY}h_{33},$$

$$g_{12} = \tilde{P}\Big\{(m^2/h)zH_{11} + H_{12} + (j/2)zH_{13}$$

$$-(mY+n)^{-1}\left[mzH_{13} + (h/m)H_{23} + \frac{1}{2}(jh/m)zh_{33}\right]\Big\},$$

$$g_{13} = H_{13}\tilde{P} - (h/m)e^{-jY/2}h_{33},$$

$$g_{22} = (m^2 z/h)^2 H_{11} + 2(m^2/h)zH_{12} + (jm^2/h)z^2 H_{13} + H_{22} + jzH_{23} + (jz/2)^2 h_{33},$$

$$g_{23} = (m^2/h)zH_{13} + H_{23} + (j/2)zh_{33}, \qquad g_{33} = h_{33} = H_{33}, \tag{12.39}$$

In the limit $\omega_0 \to 0$, the $H_{ij}$ will depend only on $t$.

The $k = -1$ Friedmann limit is now obtained when

$$H_{ij} = -C_{ij}R^2(t), \qquad h = Hm, \qquad m \to 0. \tag{12.40}$$

The $k = 0$ limit will result when $j = 0$ in addition.

# XIII. The cases 2.2.2 of Ref. 3.

The case 2.2.2.1.1 is given by eqs. (11.11) – (11.12) in Ref. 3. The transformation back to the Plebański coordinates is the inverse of (11.10), and the transformed metric is

$$g_{11} = y^2 h_{11}, \qquad g_{12} = (jt/a)(1 - h_{11}) + h_{12} - \mathcal{A}jth_{13},$$

$$g_{13} = B(j+a)y h_{13},$$

$$g_{22} = \left(\frac{jt}{ay}\right)^2 (h_{11} - 1) - 2\frac{jt}{ay^2}h_{12} + 2\mathcal{A}\frac{(jt)^2}{ay^2}h_{13} + h_{22}/y^2 - 2(\mathcal{A}jt/y^2)h_{23} + (\mathcal{A}jt/y)^2 h_{33},$$

$$g_{23} = B(j+a)\left(-\frac{jt}{ay}h_{13} + h_{23}/y - \frac{\mathcal{A}jt}{y}h_{33}\right), \qquad g_{33} = B^2(j+a)^2 h_{33}, \qquad (13.1)$$

where $a$, $\mathcal{A}$, $B$ and $j$ are arbitrary constants, and the $h_{ij}$ are arbitrary functions of the argument

$$T = t + \frac{B(j+a)}{\mathcal{A}a}z. \qquad (13.2)$$

The Killing fields for the metric (13.1) are

$$k^\alpha_{(1)} = \delta^\alpha_1, \qquad k^\alpha_{(2)} = x\delta^\alpha_1 - y\delta^\alpha_2,$$

$$k^\alpha_{(3)} = y^{-j/a}\left[B(j/a+1)\delta^\alpha_0 - Bj(ay)^{-1}\delta^\alpha_1 - \mathcal{A}\delta^\alpha_3\right], \qquad (13.3)$$

and they form a Bianchi type VI$_h$ algebra, with the free parameter $(j-a)/(j+a)$. The analog of (5.1) is

$$k^\alpha_{(3)} - Bj(ay)^{-1}k^\alpha_{(1)} = y^{-j/a}\left[B(j/a+1)u^\alpha - (\mathcal{A}/n)w^\alpha\right], \qquad (13.4)$$

and the King-Ellis measure of tilt is

$$\sqrt{-g}\,u^\alpha N_\alpha = \mathcal{A}y^{1-j/a}. \qquad (13.5)$$

By a simple transformation of the $z$-coordinate we can achieve the same result as if

$$B(j+a) = 1, \qquad (13.6)$$

and we will assume this now.

The redefinitions needed to calculate the limit $\omega_0 \to 0$ are

$$y = \omega_0 \tilde{y}, \qquad \mathcal{A} = A/\omega_0, \qquad h_{11} = H_{11}/\omega_0^2, \qquad h_{13} = H_{13}/\omega_0,$$

$$h_{12} = -(j/a)T + (jT/a)H_{11}/\omega_0^2 + H_{12}/\omega_0 + AjTH_{13}/\omega_0^2,$$

$$h_{22} = (jT/a)^2\left(1 - H_{11}/\omega_0^2\right) + 2(jT/a)h_{12} - 2(Aj^2T^2/a)H_{13}/\omega_0^2 + H_{22}$$

$$+ 2AjTh_{23}/\omega_0 - (AjT)^2 h_{33}/\omega_0^2,$$

$$h_{23} = (jT/a)H_{13}/\omega_0 + H_{23} + AjTh_{33}/\omega_0. \qquad (13.7)$$

The redefined metric is

$$g_{11} = \tilde{y}^2 H_{11}, \qquad g_{12} = \frac{j}{a^2 A} z(H_{11} - \omega_0{}^2) + H_{12} + (j/a) z H_{13}, \qquad g_{13} = \tilde{y} H_{13},$$

$$g_{22} = \tilde{y}^{-2} \left[ \left(\frac{jz}{a^2 A}\right)^2 (H_{11} - \omega_0{}^2) + 2\frac{j}{a^2 A} z H_{12} + 2\frac{j^2}{a^3 A} z^2 H_{13} + H_{22} \right.$$
$$\left. + 2(j/a) z H_{23} + (jz/a)^2 h_{33} \right],$$

$$g_{23} = \tilde{y}^{-1} \left[ \frac{j}{a^2 A} z H_{13} + H_{23} + (j/a) z h_{33} \right], \qquad g_{33} = h_{33} = H_{33}, \qquad (13.8)$$

In the limit $\omega_0 \to 0$, the $H_{ij}$ will depend only on $t$. The $k = -1$ Friedmann limit will then result when

$$H_{ij} = -C_{ij} R^2(t), \qquad j = -a, \qquad A \to \infty. \qquad (13.9)$$

The $k = 0$ Friedmann limit will result when

$$H_{ij} = -C_{ij} R^2(t), \qquad C_{13} = j = 0, \qquad C_{22} = (D_{22}/k)^2,$$

$$\tilde{y} = e^{ku}, \qquad k \to 0. \qquad (13.10)$$

The limit (13.9) transforms the Killing fields (13.3) into a Bianchi type V algebra, provided $k^\alpha_{(3)}$ is redefined to $k'^\alpha_{(3)} = -(\omega_0{}^{j/a}/\mathcal{A}) k^\alpha_{(3)}$, and the limits $A \to \infty$ and $j = -a$ are tuned so that $A(j+a) \to \infty$ (for example, $A = \alpha/(j+a)$ and $\alpha \to \infty$). The limit (13.10) will transform (13.3) into a type I algebra, but $k^\alpha_{(3)}$ has to be redefined as above, and in addition $k^\alpha_{(2)}$ has to be redefined to $k'^\alpha_{(3)} = k k^\alpha_{(2)}$.

The formulae for the case 2.2.2.1.2 simply follow from those above. This case has a completely different outlook only in the coordinates adapted to the Killing fields that were used in Ref. 3. When transformed to the Plebański coordinates, it becomes the subcase of (13.1) given by

$$B(j+a) = 1, \qquad j = -a, \qquad \mathcal{A} = \mathcal{A}_1/a, \qquad (13.11)$$

where the $\mathcal{A}_1$ defined above stands in place of the $\mathcal{A}$ from eqs. (11.17) in Ref. 3. Then the redefinitions needed are (13.7) with $\mathcal{A}_1 = A/\omega_0$ and $j = -a$, and the redefined metric is

$$g_{11} = \tilde{y}^2 H_{11}, \qquad g_{12} = -(z/A)(H_{11} - \omega_0{}^2) + H_{12} - z H_{13}, \qquad g_{13} = \tilde{y} H_{13},$$

$$g_{22} = \tilde{y}^{-2} \left[ (z/A)^2 (H_{11} - \omega_0{}^2) - 2(z/A) H_{12} + 2(z^2/A) H_{13} + H_{22} - 2z H_{23} + z^2 h_{33} \right],$$

$$g_{23} = \tilde{y}^{-1} [-(z/a) H_{13} + H_{23} - z h_{33}], \qquad g_{33} = h_{33} = H_{33}, \qquad (13.12)$$

The $k = -1$ Friedmann limit results now by (13.9) when $\omega_0 = 0$, and the $k = 0$ Friedmann limit results from (13.12) when

$$\omega_0 = 0, \qquad H_{ij} = -C_{ij} R^2(t), \qquad (\tilde{y}, z) = (e^{ku}, e^{kv}),$$

$$(H_{12}, H_{13}) = (G_{12}, G_{13})/k, \qquad (H_{22}, H_{23}, H_{33}) = (G_{22}, G_{23}, G_{33})/k^2, \qquad k \to 0. \quad (13.13)$$

The Killing field $k^\alpha_{(3)}$ is different here

$$k^\alpha_{(3)} = y \delta^\alpha{}_0 - \ln y \delta^\alpha{}_1 - \mathcal{A} y \delta^\alpha{}_3, \qquad (13.14)$$

while the two others are as in (13.3). The case 2.2.2.1.2 required a separate consideration in Ref. 3 only because of the logarithm term in the Killing field. In calculating the limits $\omega_0 \to 0$ and $A \to \infty$, this vector field has to be redefined similarly as before. For calculating the $k = 0$ Friedmann limit, $k_{(3)}^\alpha$ and $k_{(2)}^\alpha$ have to be redefined by $(k_{(2)}'^\alpha, k_{(3)}'^\alpha) = k(k_{(2)}^\alpha, k_{(3)}^\alpha) \xrightarrow[k \to 0]{} (-\delta^\alpha{}_2, -\mathcal{A}\delta^\alpha{}_3)$.

Finally, the case 2.2.2.2 from Ref. 3 (eqs. (11.18) – (11.27)) is of Bianchi type I, with the velocity field being tangent to the symmetry orbits, so it has no Friedmann limit at all.

## XIV. Summary.

All the metrics derived in Refs. 1 – 3, that correspond to rotating hypersurface-homogeneous dust models, have been checked here for the existence of a Friedmann limit. It was found that such a limit exists for all those cases listed in Refs. 2 and 3, where the matter-density is not constant along the flow. However, in at least one class (see sec. III), the Friedmann model will have no rotating parent solution, but will instead be a separate subclass.

Along the way, the nonstationary metrics were all transformed to such a form, in which the limit of zero rotation can be explicitly calculated. The transformation/reparametrization leading to this form is nonsingular and invertible in each case, but it becomes singular when $\omega \to 0$. The limits $\omega = 0$ all have nonzero shear. Thus, a whole collection of metrics generalizing those of Friedmann was found that can be used in studying spatially homogeneous exact perturbations of the latter.

The Class A Bianchi-type metrics (those in which the structure constants have the property $C^a{}_{ac} = 0$) are known to admit a lagrangian/hamiltonian formulation[12]. Those of them that obey the Einstein equations with a rotating dust source (types $VI_0$, $VII_0$, VIII and IX) were studied by Ozsváth[7,13]. The lagrangians and hamiltonians were explicitly found in Refs. 7 and 13, and the Einstein equations in the hamiltonian form were then transformed to such variables, in which they become analytic. This should prove the existence of solutions.

Two more papers, specifically devoted to rotating spatially homogeneous dust solutions, are those of Behr[11] (where a subclass of type IX models was investigated) and of this author[4] (discussing a subclass of type V models). In both of these, the Einstein equations were transformed, simplified, investigated for known limiting cases and for Lie symmetries, but no explicit solutions were found. A (hopefully) complete overview of other solutions with rotating matter source is given at the end of Ref. 3.

It is hoped that the present paper will be helpful in picking out those models for future investigation that promise interesting physics or geometry.

**Acknowledgements.** The algebraic manipulations for this paper were carried out using the computer algebra system Ortocartan[14,15].

### References.

## CAPTION TO THE DIAGRAM

The diagram shows how the different Bianchi types can be specialized by taking the zero limit of one or more of the structure constants. This allows to recognize (by the rules given at the end of sec. I), which Friedmann models can possibly be contained as limits in a given class – see text. All the possibilities are actually realized in the collection considered in the paper.

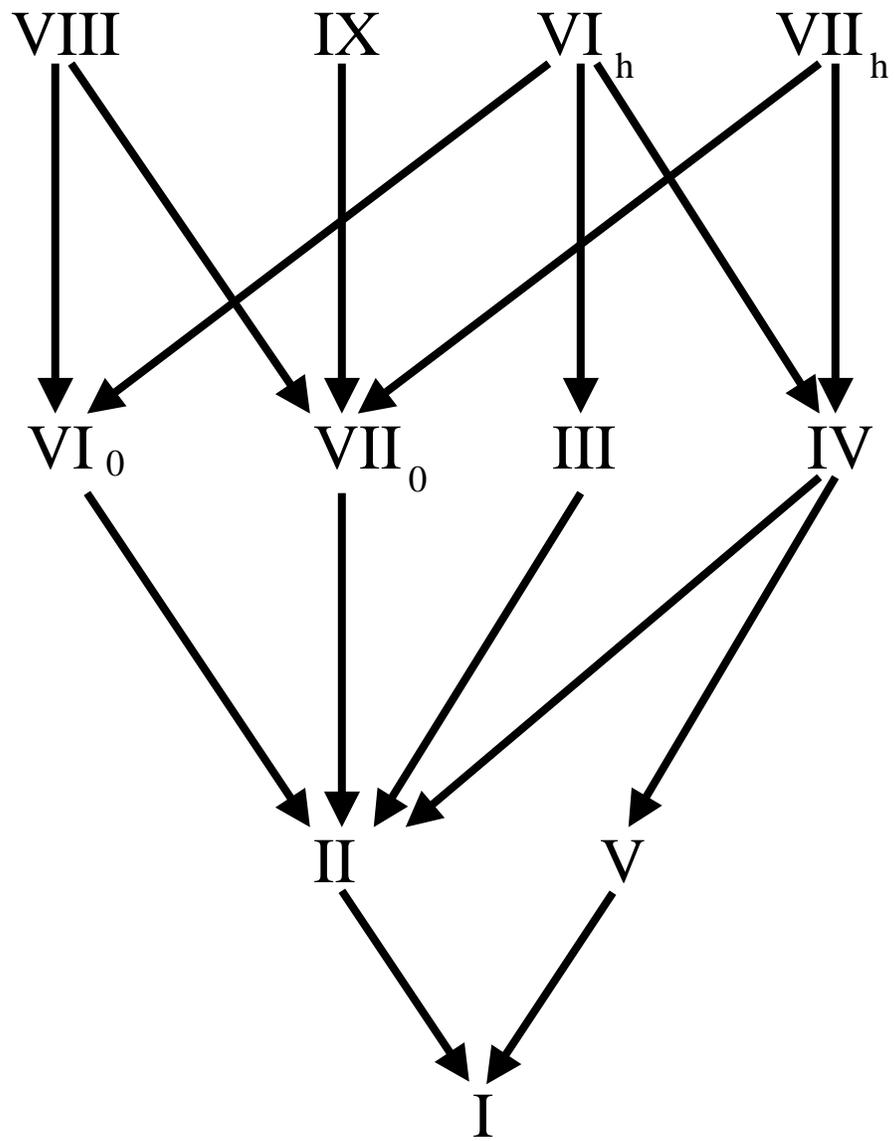